\documentclass[10pt]{article}

\usepackage[letterpaper, twocolumn, left=1.8cm, right=1.8cm, top=1.4cm, bottom=2.2cm]{geometry}
\usepackage{multicol}
\usepackage{hyphenat}
\usepackage[english]{babel}
\def\hyph{-\penalty0\hskip0pt\relax}

\usepackage[numbers,square]{natbib}
\usepackage{doi}
\usepackage[table]{xcolor}
\usepackage{hyperref}
\hypersetup{
	colorlinks,
	linkcolor={black},
	citecolor={black},
	urlcolor={blue}}

\usepackage{datetime}
\newdateformat{monthyeardate}{\monthname[\THEMONTH], \THEYEAR}

\usepackage{booktabs,tabu,multirow,array}
\usepackage{float}
\floatstyle{plaintop}
\restylefloat{table}

\newcommand{\dashedrule}{\tabucline[0.4pt black!50 off 2pt]{-}}

\usepackage{colortbl} 

\usepackage{graphicx}
\usepackage[abs]{overpic}
\renewcommand{\part}[1]{\textit{\textbf{#1.}}}
\usepackage[font=small,labelfont=bf]{caption}
\usepackage{pgfplots}
\pgfplotsset{compat=1.16}
\usepackage{tikz}
\usetikzlibrary{decorations.pathreplacing}
\usetikzlibrary{arrows.meta}
\usetikzlibrary{decorations.pathmorphing}
\usetikzlibrary{positioning}
\usetikzlibrary{calc}

\usepackage{outline}

\usepackage{amsmath, amssymb, bm, nicefrac, mathtools, xfrac}
\usepackage[version=4]{mhchem}
\DeclareMathOperator{\sgn}{sgn}
\renewcommand{\vec}[1]{\mathbf{#1}}

\newcommand{\bra}[1]{\langle#1|}
\newcommand{\ket}[1]{|#1\rangle}

\newcommand{\ee}[1]{\times10^{#1}}

\newcommand{\It}{I_\text{t}}
\newcommand{\Evec}{\bm{\mathcal{E}}}
\newcommand{\Esca}{\mathcal{E}}
\newcommand{\Bvec}{\bm{\mathcal{B}}}
\newcommand{\Bsca}{\mathcal{B}}

\newcommand{\kB}{k_\text{B}}

\newcommand{\eV}{\,\mathrm{eV}}
\newcommand{\kVcm}{\,\mathrm{kV/cm}}

\newcommand{\Vcm}{\,\mathrm{V/cm}}
\newcommand{\kV}{\,\mathrm{kV}}
\newcommand{\cm}{\,\mathrm{cm}}

\newcommand{\Hz}{\,\mathrm{Hz}}
\newcommand{\kHz}{\,\mathrm{kHz}}
\newcommand{\MHz}{\,\mathrm{MHz}}
\newcommand{\GHz}{\,\mathrm{GHz}}
\newcommand{\mHz}{\,\mathrm{mHz}}
\newcommand{\mm}{\,\mathrm{mm}}
\newcommand{\nm}{\,\mathrm{nm}}
\newcommand{\mW}{\,\mathrm{mW}}

\newcommand{\kelvin}{\,\mathrm{K}}
\newcommand{\mG}{\,\mathrm{mG}}
\newcommand{\gauss}{\,\mathrm{G}}
\newcommand{\sccm}{\,\mathrm{sccm}}

\newcommand{\CENTREX}{CeNTREX}

\renewcommand{\footnoterule}{%
  \kern -3pt
  \hrule width \columnwidth height 1pt
  \kern 5pt
}

\newcommand\blfootnote[1]{%
  \begingroup
  \renewcommand\thefootnote{}\footnote{#1}%
  \addtocounter{footnote}{-1}%
  \endgroup
}

\DeclareUnicodeCharacter{2212}{-}

\begin{document}
\renewcommand{\hbar}{h\hspace*{-0.3em}\bar{}\hspace*{0.25em}}

\twocolumn[
\begin{@twocolumnfalse}
	\begin{center}
		\begin{minipage}{.8\textwidth}
			\centering
			\vspace{4.5ex}

			{\bf\large CeNTREX: A new search for time-reversal symmetry violation in the $^{205}$Tl nucleus}

			\vspace{1.5ex}
			O. Grasdijk$^1$, O. Timgren$^1$, J. Kastelic$^1$, T. Wright$^{1,*}$, S. Lamoreaux$^1$, D. DeMille$^{2,3,1}$
			{\it\small 
			$^1$Department of Physics, Yale University, New Haven, CT 06511\\
			$^2$Physics Division, Argonne National Laboratory, Argonne, IL 60439\\
			$^3$James Franck Institute and Department of Physics, University of Chicago, Chicago, IL 60637
			}

			\vspace{1.5ex}
			K. Wenz, M. Aitken, T. Zelevinsky

			{\it\small Department of Physics, Columbia University, New York, NY 10027-5255}

			\vspace{1.5ex}
			T. Winick, D. Kawall

			{\it\small Department of Physics, University of Massachusetts Amherst, Amherst, MA 01003}
		\end{minipage}

		\vspace{3ex}

		\begin{minipage}{.75\textwidth}
			\small
			The Cold molecule Nuclear Time-Reversal EXperiment (CeNTREX) is a new effort aiming for a significant increase in sensitivity over the best present upper bounds on the strength of hadronic time reversal ($T$) violating fundamental interactions. The experimental signature will be shifts in nuclear magnetic resonance frequencies of $^{205}$Tl in electrically-polarized thallium fluoride (TlF) molecules. Here we describe the motivation for studying these $T$-violating interactions and for using TlF to do so. To achieve higher sensitivity than earlier searches for $T$-violation in TlF, CeNTREX uses a cryogenic molecular beam source, optical state preparation and detection, and modern methods of coherent quantum state manipulation. Details of the measurement scheme and the current state of the apparatus are presented, with quantitative measurements of the TlF beam. Finally, the estimated sensitivity and methods to control systematic errors are discussed. 

		\end{minipage}

		\vspace{3ex}
	\end{center}
\end{@twocolumnfalse}
]

\section{Introduction}
\label{sec:introduction}
Before the early 1960s, it was believed that $CP$ is a good symmetry of nature. As Landau pointed out, that would make it impossible for particles to have electric dipole moments (EDMs) along their spin axis \cite{landau1957conservation}. The detection of such an EDM would thus provide clear evidence of $CP$-violation (CPV).  \blfootnote{*Present address: JILA, National Institute of Standards and Technology and University of
Colorado, and Department of Physics, Boulder, Colorado 80309, USA}

While most processes preserve $CP$, certain weak interactions violate it as observed in $K$-, $B$-, and $D$-meson decays \cite{christenson1964evidence, PDG18}. The flavor-changing part of the Standard Model (SM) quark sector includes a CPV phase in the CKM quark-mixing matrix \cite{peccei1995}. This so-called Kobayashi-Maskawa mechanism introduces the third quark generation to explain the CPV \cite{kobayashi1973cp}. The CKM phase has been the only source of observed CPV so far~\cite{PDG18}.

A major motivation for CPV searches comes from the baryon asymmetry of the universe (BAU). Compared to the current baryon density $n_{\text{B}}$, the antibaryon density $n_{\overline{\text{B}}}$ is very small; the reported upper bounds for the antimatter-to-matter number ratio range from $10^{-15}$ to $10^{-6}$ \cite{canetti2012matter}. To date, no mechanism has been experimentally verified that can explain the BAU. In a 1967 paper, Sakharov argued that CPV is necessary to explain the BAU \cite{sakharov1991violation} if the initial conditions of the universe were $C$-symmetric. The existing CPV in the CKM matrix is not enough to explain the extent of the BAU \cite{pospelov2005electric}. Thus new sources of CPV are required to explain the BAU.

No flavor-neutral CPV signal has been observed yet. However, many mechanisms can lead to such phenomena. For example, the QCD Lagrangian can, in principle, include an effective CPV term, proportional to the parameter $\bar{\theta}$~\cite{pospelov1999theta}:
\begin{equation}
	\mathcal{L} = \bar{\theta}\frac{g^2}{32\pi^2}G^a_{\mu\nu}\widetilde{G}^a_{\mu\nu},
\end{equation}
where $G^a$ is the gluon field tensor, $g$ is the strong coupling constant, and $\bar{\theta}$ is dimensionless. Experimental limits from experiments searching for CPV in neutral $^{199}$Hg atoms \cite{graner2016reduced} and ultracold neutrons \cite{baker2006improved, abel2020measurement} suggests that the strength of this term relative to the usual strong interaction is $\left|\bar{\theta}\right|<9\cdot10^{-11}$. The unexplained smallness of $\bar{\theta}$ is known as the strong $CP$ problem. One proposed solution to the strong $CP$ problem is the so called Peccei-Quinn mechanism, with an accompanying elementary scalar particle: the axion \cite{PhysRevLett.38.1440}. The axion would naturally lead to $\bar{\theta}\approx 0$, and is an attractive candidate for dark matter~\cite{PRESKILL1983127,PhysRevLett.50.925,PhysRevLett.124.101303}. (A review of experimental searches for the axion is given in \cite{graham2015experimental}.)

New hadronic $CP$-violating interactions from the QCD sector, or from physics beyond the SM, can lead to an effective charge asymmetry along the spin of a particle. Such charge asymmetries include EDMs and, for finite size particles such as nuclei, Schiff moments \cite{schiff1963measurability}. In the Standard Model, EDMs and nuclear Schiff moments (NSMs) are induced by the CKM phase, but are strongly suppressed: an EDM cannot appear below the three-loop level for quarks, or four-loop for leptons \cite{pospelov1991electric}. The CKM phase can produce a proton or neutron EDM no larger than $10^{-32}\,e\,$cm. Proposed extensions to the SM carry new CPV phases, which may manifest as EDMs or NSMs larger than expected based on the Standard Model. The search for an EDM or NSM thus constitutes a nearly background-free signal for new physics. In fact, the background expected from the Standard Model would only become apparent when probing effects beyond the energy scale of $\sim\!10^{5}\,$TeV \cite{pospelov2005electric}.

At present, searches for EDMs and related phenomena give the most sensitive constraints on flavor-neutral CPV effects beyond the SM. However, these searches are subject to the following limitation. According to the Schiff theorem, the interaction energy of nonrelativistic point-charged electric dipoles, bound in a neutral system but subject to an arbitrary external electrostatic potential, has no term linear in the CPV charge distribution \cite{schiff1963measurability}. Physically, the system rearranges itself so as to screen the external field completely \cite{safronova2018search}. Thus, a $CP$-violating moment of a charged constituent in a bound system cannot be detected without some mechanism to bypass Schiff's theorem. Two such mechanisms are relativistic constituent motion and finite constituent size. 

A nucleus in an atom or molecule is nonrelativistic, but has an extended size. This finite size can lead to a residual electromagnetic moment, the Schiff moment $\vec S$, that gives rise to a $CP$-violating interaction. In heavy diamagnetic atoms and diatomic molecules such as TlF, this finite-size effect gives the dominant contribution to CPV signals. Since the nuclear spin $\vec{I}$ is the only preferred direction in a nucleus, $\vec S$ has to lie parallel to this axis, i.e., $\vec{S} = S\vec{I}/I$. This quantum Schiff moment has the symmetries of $\vec{I}$: it changes sign under time reversal ($T$) but not under parity ($P$). By contrast, the classical Schiff moment is a static charge distribution that is unchanged under $T$ but changes sign under $P$. Hence, a nonzero value of $S$ means that both $T$ and $P$ symmetries are violated.
On the assumption that $CPT$ is a good symmetry, a nonzero value of $\vec S$ thus is also a signature of CPV.

The Schiff moment corresponds to a charge displacement that is similar to an EDM in its asymmetric distribution along the spin axis. It is equivalent to a charge density on the nuclear surface proportional to $\cos{\theta}$, where $\theta$ is the angle from $\vec{I}$; this surface charge distribution produces a uniform electric field inside the nucleus \cite{GingesFlambaum2004}. The magnitude $S$ of the NSM scales with the atomic mass $A$ as $S\propto A^{2/3}$ \cite{khriplovich1997}.

The value of $S$ can be related to more fundamental $CP$-violating parameters, including CPV $\pi$ meson--nucleon interaction constants $\bar{g}_{0},\,\bar{g}_1,$ and $\bar{g}_2$; the $\bar{\theta}$ QCD parameter; quark chromo-EDMs $\widetilde{d}_\text{d}$ and $\widetilde{d}_\text{u}$; and the neutron and proton EDMs, $d_\text{n}$ and $d_\text{p}$.  For example, the NSM of the $^{205}\mathrm{Tl}$ nucleus\footnote{The $^{205}\mathrm{Tl}$ nucleus has closed neutron shells; hence its NSM has negligible contribution from $d_\text{n}$.}  can be written as \cite{PhysRevA.101.042504,PhysRevA.101.042501}:
\begin{equation} 
    \label{eq:schiff_contributions}
        \begin{split}
            S\left(^{205}\mathrm{Tl}\right) & \approx \left(0.13g\bar{g}_0 - 0.004g\bar{g}_1 - 0.27 g \bar{g}_2\right) \,e\,\mathrm{fm}^3; \\
            S\left(^{205}\mathrm{Tl}\right) & \approx 0.027 \bar{\theta}\ e\,\mathrm{fm}^3; \\
            S\left(^{205}\mathrm{Tl}\right) & \approx \left(12\widetilde{d}_\text{d}+9\widetilde{d}_\text{u}\right) e\,\mathrm{fm}^2; \\
            S\left(^{205}\mathrm{Tl}\right) & \approx 0.4\,d_\text{p} \,\mathrm{fm}^2.
    \end{split}
\end{equation}
If detected, a nonzero $S\left(^{205}\mathrm{Tl}\right)$ would provide evidence for a nonzero value of one or more of these fundamental CPV parameters.

Energy shifts associated with a NSM can be greatly enhanced in polar molecules, where there is another intrinsic direction in addition to the nuclear spin: the internuclear axis $\hat{\vec{n}}$. In TlF,  we define $\hat{\vec{n}}$ as pointing from F to Tl, associated with the internal molecular dipole moment and a corresponding strong intramolecular gradient of the electron density.  For nuclei inside a molecule, the NSM (and other $CP$-violating effects \cite{khriplovich1997}) interacts with this density gradient, giving rise to an effective CPV Hamiltonian of the form \cite{PhysRevA.101.042504}
\begin{equation}\label{eq:Hamiltonian_effective_interaction}
    \mathcal{H}_\text{CPV}= W_S S\,\frac{\vec{I}}{I} \cdot \hat{\vec{n}}.
\end{equation}
Here, $W_S$ is the proportionality constant between $S$ and the CPV contribution to the molecular energy, for a fully polarized molecule. Its value is determined by the properties of the electronic wavefunctions, which can be calculated from first principles \cite{PhysRevA.101.042504, doi:10.1002/qua.20418, PhysRevLett.88.073001, doi:10.1080/00268976.2020.1767814}. The magnitude of $W_S$ grows rapidly with atomic number $Z$ of the nucleus, as $W_S\propto Z^2$ \cite{khriplovich1991parity}.

Without an external electric field $\Evec$, the interaction of Eq.~\ref{eq:Hamiltonian_effective_interaction} fails to produce a first-order effect in any given energy eigenstate. This is because the rotation of the molecule averages $\hat{\vec n}$ to zero, and the expectation value $\langle\mathcal{H}_\text{CPV}\rangle$ vanishes \cite{cho1989tenfold, cho1991search}. However, when an external field is applied, the molecule becomes polarized and both
$\hat{\vec{n}}$ and $\mathcal{H}_\text{CPV}$  acquire non-zero expectation values, with $\langle\hat{\vec n}\rangle \parallel \Evec$. We define the degree of electrical polarization $\mathcal{P} \equiv \langle \hat{\vec{n}}\cdot\hat{\Evec} \rangle$, (where $\hat{\Evec} = \Evec/\Esca$), so that $-1\leq\mathcal{P}\leq1$. 
Hence, energy shifts due to CPV are given by  $\langle\mathcal{H}_\text{CPV}\rangle = W_S S \mathcal{P} \hat{\Evec} \cdot \frac{\vec{I}}{I}$.

\begin{figure}
    \centering
    \def\svgwidth{0.5\textwidth}
    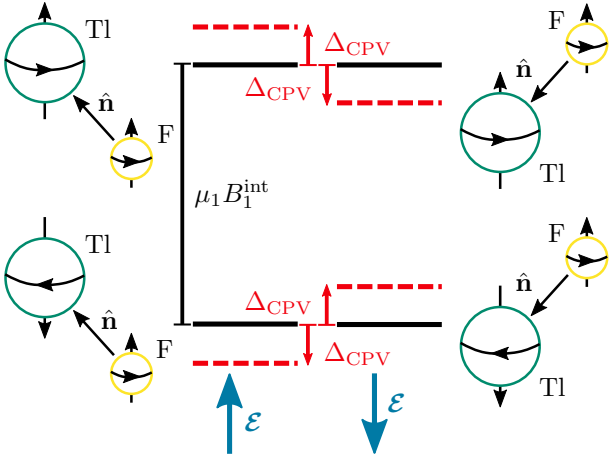
	\caption{T-violating energy shift $\Delta_{\rm CPV} = W_SS\,\mathcal{P}$ as a result of a nonzero NSM $S$ given by the effective interaction $\mathcal{H}_\text{CPV}= W_S S\vec{I}\cdot \hat{\vec{n}}/I$ (Eq.~\ref{eq:Hamiltonian_effective_interaction}), shown for opposite orientations of the applied field $\Evec$. Here $\mu_1$ is the Tl magnetic moment and $B_1^\mathrm{int}$ is the effective internal magnetic field at the Tl nucleus due to the spin-rotation interaction.
	}
	\label{fig:edm_shift}
\end{figure}

Polar molecules can be polarized readily in laboratory-scale fields owing to their small rotational level separation ($\sim\!10^{-4}\eV$), giving them near-maximal energy shifts induced by a given Schiff moment. Thus, Sandars \cite{sandars1967measurability} suggested a molecular-beam resonance experiment could be used to probe the existence of the proton EDM, if the molecule has a heavy atom with an unpaired proton in the nucleus, such as $^{205}\mathrm{Tl}$. The value of $S$ is determined by measuring energy splittings between spin-up and -down states relative to $\left\langle \hat{\vec{n}}\right\rangle$ (which is parallel to the applied electric field $\bm{\mathcal{E}}$). This splitting will increase or decrease as $\bm{\mathcal{E}}$ (and hence $\left\langle \hat{\vec{n}}\right\rangle$) reverses, due to the interaction in Eq.~\ref{eq:Hamiltonian_effective_interaction}. The difference in level splittings is proportional to the electric polarization $\mathcal{P}$, the interaction strength $W_S$, and $S$ (Fig. \ref{fig:edm_shift}).

\CENTREX\ uses a cold beam of thallium fluoride (TlF) to measure nuclear $T$-violation due to the NSM of the $^{205}$Tl nucleus. It is a suitable system to look for $P$- and $T$-violating interactions for a number of reasons: a molecular beam of thallium fluoride can be readily obtained; many of the molecular states and transitions are known experimentally; the species is a polar diatomic molecule, enhancing the electron density gradient at the site of the nuclei (and hence $W_S$). As the thallium nucleus is heavy $\left(A=205,\,Z=81\right)$ and the NSM-induced energy shift scales $\propto A^{2/3}Z^2$, the observable effect of the Tl Schiff moment is correspondingly large \cite{sandars1967measurability, wilkening1984search}. Since the Tl nucleus contains an unpaired proton,
\CENTREX\ will be primarily sensitive to proton EDM effects, as opposed to other leading experiments which are more sensitive to the
neutron EDM \cite{graner2016reduced}. TlF is not very sensitive to the electron EDM due to its zero total electron spin \cite{kozlov1995parity}. 

The current best constraint on $T$-violating interactions associated with $S\left(^{205}\mathrm{Tl}\right)$ was found by Cho, Sangster and Hinds in 1991 \cite{cho1989tenfold, cho1991search}, who measured a NSM-induced frequency shift of $\Delta E = 2\Delta_{\rm CPV} = \left(1.4\pm 2.4\right)\times 10^{-4}$~Hz, consistent with zero.\footnote{Throughout, we express both frequencies and energies in linear frequency units (Hz), and all angular momentum operators are treated as dimensionless.} 
Using the effective interaction $\mathcal{H}_\text{CPV}$, the shift in the energy splitting between states with Tl spin up versus spin down, relative to the quantization axis, can be interpreted as
\begin{equation}
    \Delta E= 2 \Delta_{\rm CPV} = 2 W_S\,S\, \textrm{sgn}(\Esca) \,\mathcal{P},
    \label{eq:frequency_shift_due_to_NSM}
\end{equation}
where $W_S = 40539\,$a.u.,
polarization $\mathcal{P} = \langle\hat{\vec{n}} \cdot \hat{\Evec}\rangle = 0.547$, and the sign of $\Esca$ refers to the direction of $\Evec$ relative to a fixed quantizing axis $\hat{z}$.  This determines the Schiff moment \cite{PhysRevA.101.042501, PhysRevLett.88.073001}
\begin{equation}
    S\left(^{205}\mathrm{Tl}\right) = \left(3.6\pm 6.1\right)\times 10^{-11}\,e\,\mathrm{fm}^3.
\end{equation}
With Eq.~\ref{eq:schiff_contributions}, the following limits can be placed:
\begin{equation}
    \label{eq:prev_best_lims}
    \begin{split}
        \bar{\theta} & = \left(1.3 \pm 2.3\right)\ee{-9}, \\
        12\bar{d}_d+9\bar{d}_u & = \left(3.6\pm 6.1\right)\ee{-24}\,\mathrm{cm}, \\
        d_p          & = \left(0.9 \pm 1.5\right)\ee{-23}\,e\,\mathrm{cm},\\
        0.13g\bar{g}_0 - 0.004g\bar{g}_1-0.27g\bar{g}_2 & = \left(3.6 \pm 6.1\right)\ee{-11}.
    \end{split}
\end{equation}
\CENTREX\ aims to improve on these limits by using a cryogenic molecular beam source to achieve a cold beam with higher intensity and lower velocity spread compared to the jet source used in the previous work. Rotational cooling will be performed with optical and microwave pumping, collapsing much of the initial Boltzmann distribution into one state, greatly enhancing the number of molecules accessible for measurement. Finally, optical cycling will be used to assist state readout, resulting in near-unity detection efficiency. Fluorescence detection, compared to the hot-wire techniques used previously, allows for background-free detection if scattered light is well controlled.

\subsection{Thallium Fluoride}
\label{sec:tlf_theory}

\begin{figure}
    \usetikzlibrary{decorations.markings}
    \def\MarkLt{4pt}
    \def\MarkSep{2pt}
    \tikzset{
      TwoMarks/.style={
        postaction={decorate,
          decoration={
            markings,
            mark=at position #1 with
              {
                  \begin{scope}[xslant=0.2]
                  \draw[line width=\MarkSep,white,-] (0pt,-\MarkLt) -- (0pt,\MarkLt) ;
                  \draw[-] (-0.5*\MarkSep,-\MarkLt) -- (-0.5*\MarkSep,\MarkLt) ;
                  \draw[-] (0.5*\MarkSep,-\MarkLt) -- (0.5*\MarkSep,\MarkLt) ;
                  \end{scope}
              }
           }
        }
      },
      TwoMarks/.default={0.5},
    }

\begin{tikzpicture}[scale=5]

   \def\len{.05}

   \def\pos{-18}
   \def\lenmark{.025}
   \def\xoffset{.18}

   \def\xa{-4.5*\len}
   \def\xb{-2.5*\len}
   \def\xc{-0.5*\len}
   \def\xd{+1.5*\len}
   \def\xe{+3.5*\len}
   \def\xf{-6.5*\len}
   \def\xg{+5.5*\len}

   \def\yA{.6}
   \def\yB{\yA+.7}

   \draw[->,TwoMarks=0.2,TwoMarks=0.6] (\xf-\len-\xoffset, -.1)
      -- (\xf-\len-\xoffset,\yB+.25) node[above, xshift = -20] {$E$ [kHz]};

   \def\bracePos{\xg+3*\len}

   \draw [decorate,decoration={brace,amplitude=5pt,mirror,raise=4pt},yshift=0pt]
      (\bracePos,-.08) -- node (midJ0) {} (\bracePos,.08) node [black,midway,anchor=west,xshift=25] (nodeJ0) {$0$};
   \draw (\xf-\len-0.5*\lenmark-\xoffset, -.003325) node[anchor=east,below,xshift=\pos]
      {-3.325} -- (\xf-\len+0.5*\lenmark-\xoffset,-.003325) node[right,below,xshift=-\pos] {$1$};
   \draw (\xf-\len-0.5*\lenmark-\xoffset, .009975) node[anchor=east,above,xshift=\pos]
      {9.975} -- (\xf-\len+0.5*\lenmark-\xoffset,.009975) node[right,above,xshift=-\pos] {$0$};

   \draw (\xc, -0.003325) -- (\xc+\len, -0.003325);
   \draw (\xb, -0.003325) -- (\xb+\len, -0.003325);
   \draw (\xd, -0.003325) -- (\xd+\len, -0.003325);
   \draw (\xc, 0.009975) -- (\xc+\len, 0.009975);

   \draw [decorate,decoration={brace,amplitude=10pt,mirror,raise=4pt},yshift=0pt]
      (\bracePos,\yA-.2) -- node (midJ1) {} (\bracePos,\yA+.1) node [black,midway,anchor=west,xshift=25] (nodeJ1) {$1$};
   \draw (\xf-\len-0.5*\lenmark-\xoffset, \yA-.143745) node[anchor=east,below,xshift=\pos] {-143.745} --
      (\xf-\len+0.5*\lenmark-\xoffset,\yA-.143745) node[right,below,xshift=-\pos] {$0$};
   \draw (\xf-\len-0.5*\lenmark-\xoffset, \yA-.121506) node[anchor=east,above,xshift=\pos] {-121.506} --
      (\xf-\len+0.5*\lenmark-\xoffset,\yA-.121506) node[right,above,xshift=-\pos] {$1$};
   \draw (\xf-\len-0.5*\lenmark-\xoffset, \yA+.054446) node[anchor=east,below,xshift=\pos] {54.446} --
      (\xf-\len+0.5*\lenmark-\xoffset,\yA+.054446) node[right,below,xshift=-\pos] {$1$};
   \draw (\xf-\len-0.5*\lenmark-\xoffset, \yA+.068985) node[anchor=east,above,xshift=\pos] {68.985} --
      (\xf-\len+0.5*\lenmark-\xoffset,\yA+.068985) node[right,above,xshift=-\pos] {$2$};

   \draw (\xc, \yA-.143745) -- (\xc+\len, \yA-.143745);
   \draw (\xc, \yA-.121506) -- (\xc+\len, \yA-.121506);
   \draw (\xb, \yA-.121506) -- (\xb+\len, \yA-.121506);
   \draw (\xd, \yA-.121506) -- (\xd+\len, \yA-.121506);
   \draw (\xc, \yA+.054446) -- (\xc+\len, \yA+.054446);
   \draw (\xb, \yA+.054446) -- (\xb+\len, \yA+.054446);
   \draw (\xd, \yA+.054446) -- (\xd+\len, \yA+.054446);
   \draw (\xc, \yA+.068985) -- (\xc+\len, \yA+.068985);
   \draw (\xb, \yA+.068985) -- (\xb+\len, \yA+.068985);
   \draw (\xd, \yA+.068985) -- (\xd+\len, \yA+.068985);
   \draw (\xa, \yA+.068985) -- (\xa+\len, \yA+.068985);
   \draw (\xe, \yA+.068985) -- (\xe+\len, \yA+.068985);

   \draw [decorate,decoration={brace,amplitude=10pt,mirror,raise=4pt},yshift=0pt]
      (\bracePos,\yB-.27) -- node (midJ2) {} (\bracePos,\yB+.18) node [black,midway,anchor=west,xshift=25] (nodeJ2) {$2$};
   \draw (\xf-\len-0.5*\lenmark-\xoffset, \yB-.217455)
   node[anchor=east,left,yshift=-5] {-217.455}
      -- (\xf-\len+0.5*\lenmark-\xoffset,\yB-.217455) node[right,below,xshift=-\pos] {$1$};
   \draw (\xf-\len-0.5*\lenmark-\xoffset, \yB-.172936)
   node[anchor=east,left,yshift=5] {-172.936}
      -- (\xf-\len+0.5*\lenmark-\xoffset,\yB-.172936) node[right,above,xshift=-\pos] {$2$};
   \draw (\xf-\len-0.5*\lenmark-\xoffset, \yB+.105876)
   node[anchor=east,left,yshift=-5] {105.876}
      -- (\xf-\len+0.5*\lenmark-\xoffset,\yB+.105876) node[right,below,xshift=-\pos] {$2$};
   \draw (\xf-\len-0.5*\lenmark-\xoffset, \yB+.141095)
   node[anchor=east,left,yshift=5] {141.095}
      -- (\xf-\len+0.5*\lenmark-\xoffset,\yB+.141095) node[right,above,xshift=-\pos] (J2F3label) {$3$};

   \draw (\xc, \yB-.217455) -- (\xc+\len, \yB-.217455);
   \draw (\xb, \yB-.217455) -- (\xb+\len, \yB-.217455);
   \draw (\xd, \yB-.217455) -- (\xd+\len, \yB-.217455);
   \draw (\xc, \yB-.172936) -- (\xc+\len, \yB-.172936);
   \draw (\xb, \yB-.172936) -- (\xb+\len, \yB-.172936);
   \draw (\xd, \yB-.172936) -- (\xd+\len, \yB-.172936);
   \draw (\xa, \yB-.172936) -- (\xa+\len, \yB-.172936);
   \draw (\xe, \yB-.172936) -- (\xe+\len, \yB-.172936);
   \draw (\xc, \yB+.105876) -- (\xc+\len, \yB+.105876);
   \draw (\xb, \yB+.105876) -- (\xb+\len, \yB+.105876);
   \draw (\xd, \yB+.105876) -- (\xd+\len, \yB+.105876);
   \draw (\xa, \yB+.105876) -- (\xa+\len, \yB+.105876);
   \draw (\xe, \yB+.105876) -- (\xe+\len, \yB+.105876);
   \draw (\xc, \yB+.141095) -- (\xc+\len, \yB+.141095);
   \draw (\xb, \yB+.141095) -- (\xb+\len, \yB+.141095);
   \draw (\xd, \yB+.141095) -- (\xd+\len, \yB+.141095);
   \draw (\xa, \yB+.141095) -- (\xa+\len, \yB+.141095);
   \draw (\xe, \yB+.141095) -- (\xe+\len, \yB+.141095);
   \draw (\xf, \yB+.141095) -- (\xf+\len, \yB+.141095);
   \draw (\xg, \yB+.141095) -- (\xg+\len, \yB+.141095);

   \draw[Latex-Latex] (nodeJ0) -- node[above,sloped] {13.3 GHz} (nodeJ1);
   \draw[Latex-Latex] (nodeJ1) -- node[above,sloped] {26.7 GHz} (nodeJ2);

   \def\FposX{-5}
   \def\FposY{15}
   \def\FposYY{22}
   \node[xshift=\FposX] at (midJ0) {\nicefrac{1}{2}};
   \node[xshift=\FposX,yshift=-\FposY] at (midJ1) {\nicefrac{1}{2}};
   \node[xshift=\FposX, yshift=\FposY] at (midJ1) {\nicefrac{3}{2}};
   \node[xshift=\FposX,yshift=-\FposYY] at (midJ2) {\nicefrac{3}{2}};
   \node[xshift=\FposX, yshift=\FposYY] at (midJ2) {\nicefrac{5}{2}};
   \node[xshift=\FposX, yshift=47] at (midJ2) (F1label) {$F_1$};
    
   \node at (F1label -| nodeJ2) {$J$};
   \node at (F1label -| J2F3label) {$F$};

\end{tikzpicture}
	\caption{Hyperfine structure in the lowest three rotational levels in the TlF ground state $X^1\Sigma^+$, with no applied fields. Rotational energies $hBJ(J+1)$ should be added to
	the hyperfine energy shifts indicated on the axis. Note the $(2F+1)$-fold degeneracy of each state with total angular momentum $F$, corresponding to the quantum numbers $m_F=-F,\dots,0,\dots,F$.}
	\label{fig:level_diagram}

\end{figure}
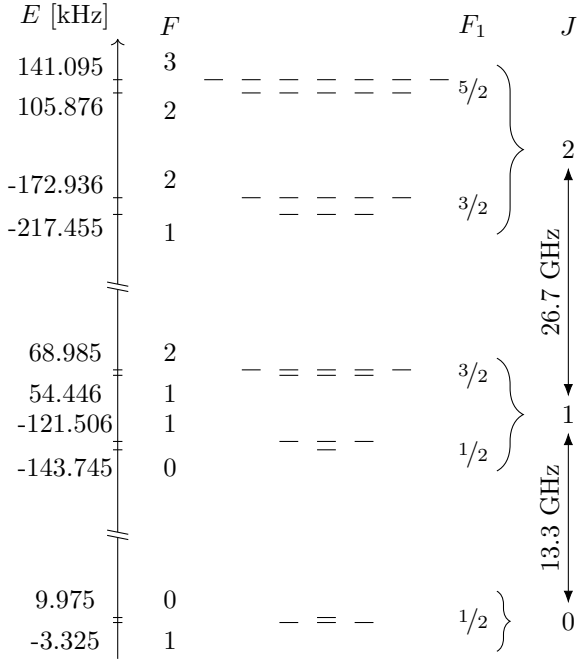

The TlF molecule is described by its electronic, vibrational and rotational motion, plus the states of the Tl and F nuclear spins. \CENTREX\ makes use of states both in the vibronic ground state, $X ^1\Sigma^+\left(\nu=0\right)$, and in an electronically excited state, $B ^3\Pi_1\left(\nu=0\right)$.  In both cases, we describe the angular momentum couplings in a Hund's case (a) basis. We typically write energy eigenstates in terms of the basis states $\ket{\eta,J,F_1,F,m_F}$. Here, $\eta$ represents the vibronic quantum numbers; $J$ is the total angular momentum excluding the nuclear spins, $\vec{F}_1 = \vec{J}+\vec{I}_\mathrm{1}$, with $I_\mathrm{1}=1/2$ for $^{205}$Tl; $\vec{F} = \vec{F}_1+\vec{I}_\mathrm{2}$ is the total angular momentum, with $I_\mathrm{2}=1/2$ for $^{19}$F; and $m_F$ is its projection along a quantization axis in the lab frame. Field-free eigenstates are close to these basis states in the ground $X ^1\Sigma^+$ state.  In the $B ^3\Pi_1$ state, strong hyperfine interactions significantly mix states with different $J$ and $F_1$ values. Hence, we describe these eigenstates with the modified notation $\ket{\eta',\widetilde{J}' ,\widetilde{F}_1^\prime,F',m_F'}$, where $\widetilde{J}'$ and $\widetilde{F}_1^\prime$ correspond to the largest component in their basis-state decomposition; the primes indicate that the ket refers to the excited state $B ^3\Pi_1$.

Molecules in the beam are assumed to be in the vibronic ground state, since the beam temperature is much lower than the energy scales associated with the electronic and vibrational excitations. However, even at cryogenic temperatures, there is a Boltzmann distribution over many rotational and nuclear spin states. The dominant term in the energy of rotational/spin levels in the $X ^1\Sigma^+$ state is due to rotation; the mean energy of states with quantum number $J$ is $E_\mathrm{rot}=BJ(J+1)$, where $B\approx 6.67\GHz \approx 0.3\kelvin\,\kB$, where $\kB$ the Boltzmann constant.\footnote{In Ramsey et al.\ \cite{wilkening1984search}, the symbol $B$ denotes the rotational constant in equilibrium position, i.e., there $B\equiv B_\text{e}$. However, the effective $v=0$ rotational constant, $B_0$, is more relevant to \CENTREX. To first order in the Dunham expansion \cite{huber2013molecular}, $B_0=B_\text{e}-\alpha_\text{e}/2$. With $B_\text{e}$ and $\alpha_\text{e}$ from the NIST database \cite{afeefy2011nist}, we find $B_0$. We define the symbol $B\equiv B_0$; its value is shown in Table~\ref{tab:hyperfine_hamiltonian}.}

Hyperfine interactions split sublevels with different $F$ values ($F=J-1,J,J, J+1$, except $F=0,1$ only for $J=0$) in each rotational state. Thus, each rotational level has $4(2J+1)$ magnetic sublevels. 
Including rotation, spin-rotation and spin-spin interactions, plus interactions with external electric $(\Evec)$ and magnetic $(\Bvec)$ fields, the system is described by the effective Hamiltonian \cite{wilkening1984search}
\small
$\mathcal{H}_\text{TlF} = \mathcal{H}_\text{rot}+\mathcal{H}_\text{sr}+\mathcal H_\text{ss} + \mathcal H_\text{S}+\mathcal H_\text{Z}$, 
\normalsize
where
\begin{equation} 
    \label{eq:hyperfine_hamiltonian}
    \begin{split}
        \mathcal H_\text{rot} & = B \vec{J}^2 \\
        \mathcal H_\text{sr}&=c_1(\vec I_1\cdot\vec J)+c_2(\vec I_2\cdot\vec J), \\
        \mathcal H_\text{ss}&=c_3 T^2(\vec C)\cdot T^2(\vec I_1, \vec I_2)+c_4(\vec I_1\cdot\vec I_2), \\
        \mathcal H_\text{S}&=-\vec\mu_e\cdot\vec \Evec,\\
        \mathcal H_\text{Z}&=-\frac{\mu_J}{J}(\vec J\cdot\vec \Bvec)-\frac{\mu_1}{I_1}(\vec I_1\cdot\vec \Bvec)-\frac{\mu_2}{I_2}(\vec I_2\cdot\vec \Bvec).
    \end{split}
\end{equation}
Here, the first term in the spin-spin interaction ($\mathcal H_\text{ss}$) contains the scalar product of two rank-2 tensors: one constructed from the modified spherical harmonics $\vec C$, and one from $\vec{I}_1$ and $\vec{I}_2$ \cite{brown2003rotational}. (The matrix elements diagonal in $J$ of this term are given in \cite{wilkening1984search}.)  The hyperfine parameters $c_1,c_2,c_3,c_4$, rotational constant $B$, magnetic moments $\mu_J,\mu_1,\mu_2$, and molecule-frame electric dipole moment $\mu_e$ are all known from previous measurements; their values are given in  Table~\ref{tab:hyperfine_hamiltonian}. A level diagram of low-lying states in the absence of applied fields is shown in Fig.~\ref{fig:level_diagram}.

\begin{table}
    \small
    \setlength\extrarowheight{3pt}
    \rowcolors{2}{white}{gray!25}
	\centering
	\caption{Constants describing rotational, hyperfine, Zeeman, and Stark interactions in the effective TlF ground-state  Hamiltonian (Eq.~\ref{eq:hyperfine_hamiltonian}). All values taken from Ramsey et al.\ \cite{wilkening1984search}, except for $B$ (see Sec. \ref{sec:tlf_theory}).}
	\label{tab:hyperfine_hamiltonian}
	\begin{tabular}{r@{\hspace{5pt}}c@{\hspace{5pt}}l@{\hspace{5pt}}l | r@{\hspace{5pt}}c@{\hspace{5pt}}l@{\hspace{5pt}}l}
    	\toprule
		$B$ & = &           $6.66733$ &     GHz &       $\mu_e$ & = & $4.2282(8)$ &     Debye \\
		$\mu_J$ & = &       $35(15)$ &      Hz/G &      $c_1$ & = & $126.03(12)$ &    kHz \\
		$\mu_1^{205}$ & = & $1.2405(3)$ &   kHz/G &     $c_2$ & = & $17.89(15)$ &     kHz\\
		$\mu_1^{203}$ & = &  $1.2285(3)$ &   kHz/G &     $c_3$ & = & $0.70(3)$ &       kHz \\
		$\mu_2$ & = &      $2.00363(4)$ &  kHz/G  &    $c_4$ & = & $-13.30(72)$ &    kHz \\
		\bottomrule
	\end{tabular}
\end{table}

In \CENTREX, lasers are tuned to $X^1\Sigma^+(\nu=0)\rightarrow B^3\Pi_1\left(\nu=0\right)$ transitions in order to manipulate and read out ground state hyperfine and rotational sublevels. Details of the $B ^3\Pi_1$ state structure are given in \cite{norrgard2017hyperfine,PhysRevA.101.042506}.  Here, only a few main features of the $B$ state substructure are important.  First, the $B$ state hyperfine splittings are very large ($\gtrsim 100\MHz$) compared to the natural linewidth of the transition ($\gamma_B \approx 1.6\MHz$), which is in turn much larger than the ground-state hyperfine splittings ($c_j \lesssim 100\kHz$). This means that hyperfine structure is fully resolved in the excited state, and entirely unresolved in the ground state. Hence, optical transitions in TlF drive a large manifold of ground-state hyperfine levels (with a given value of $J$) to a single hyperfine state with (nominal) quantum numbers $\tilde{J},\tilde{F}_1$ and exact quantum number $F$. Another important feature of the $B$ state is that its matrix of Franck-Condon factors (FCFs) for decay to the $X$ state is extremely diagonal \cite{hunter2012prospects}, such that $\sim\! 99\%$ of the time, the $B(v=0)$ vibronic state decays back to the vibronic ground state $X(v=0)$.  This enables optical pumping and optical cycling with little loss.  However, the mixing of $J$ and $F_1$ by the strong $B$-state hyperfine interaction substantially modifies rotational selection rules in $B-X$ decays, and must be taken into account when describing optical excitation and emission in TlF.

\subsection{TlF in \texorpdfstring{$\Esca$}{E}-fields}
\label{sec:TlF_in_E_fields}

\begin{figure*}
	\centering
	\includegraphics[width=\textwidth]{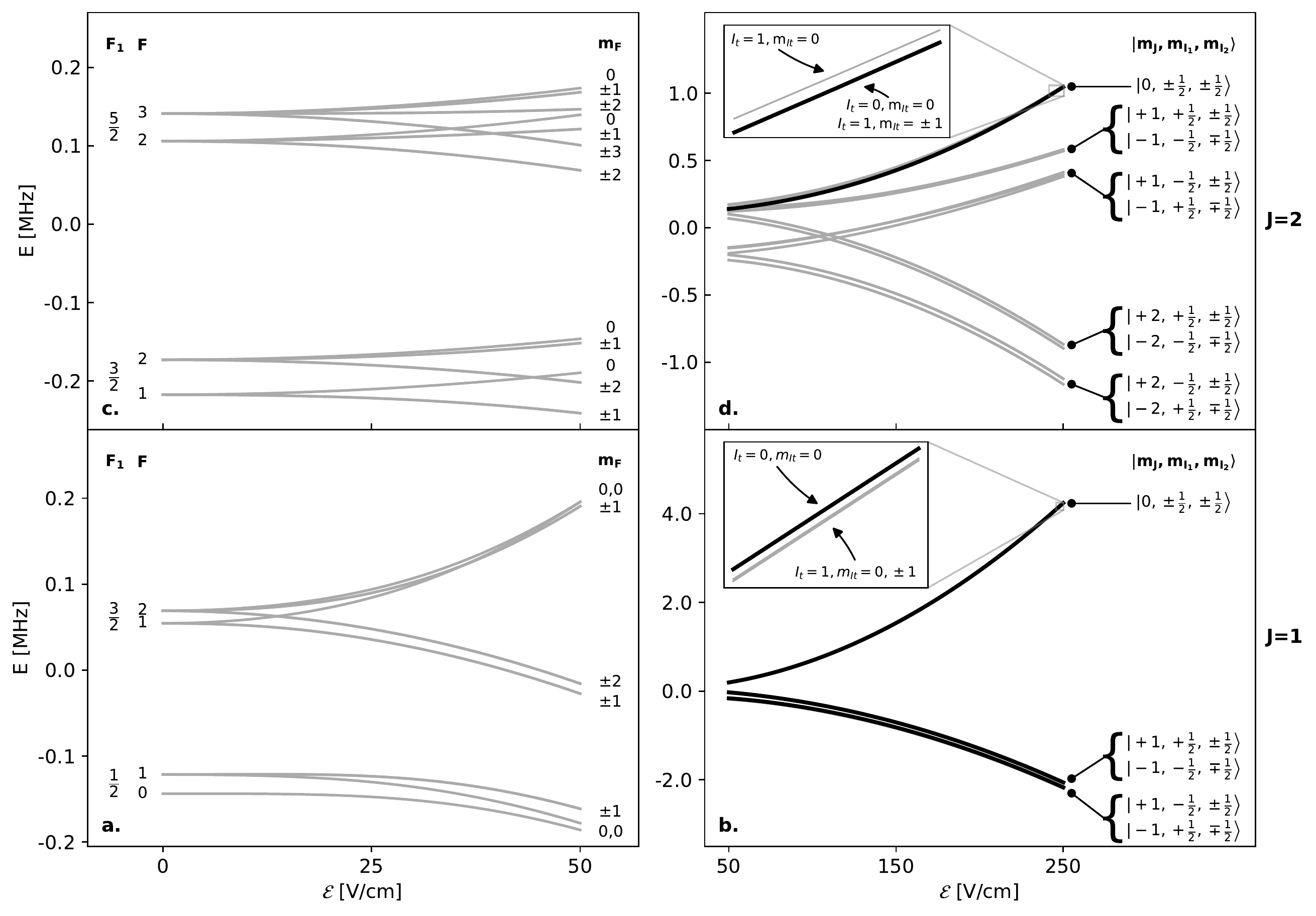}
	\caption{Overview of the energy eigenstates for changing $\Esca$-field magnitudes. The low-field regime, where $\Delta E_{\rm S} \ll E_{\rm hf}$, where energy eigenstates retain $J$, $F$, and $F_1$ as approximate quantum numbers is shown in \part{a} for $J=1$ and \part{c} for $J=2$. The mid-field regime, where $E_{\rm hf} \ll \Delta E_{\rm S} \ll E_{\rm rot}$, where both $J$ and $m_J$ are approximate quantum numbers is shown in \part{b} for $J=1$ and \part{d} for $J=2$. States used in \CENTREX\ are shown in bold.}
	\label{fig:low_to_mid_field}
\end{figure*}

\begin{figure}
	\centering
	\includegraphics[width=\textwidth/2]{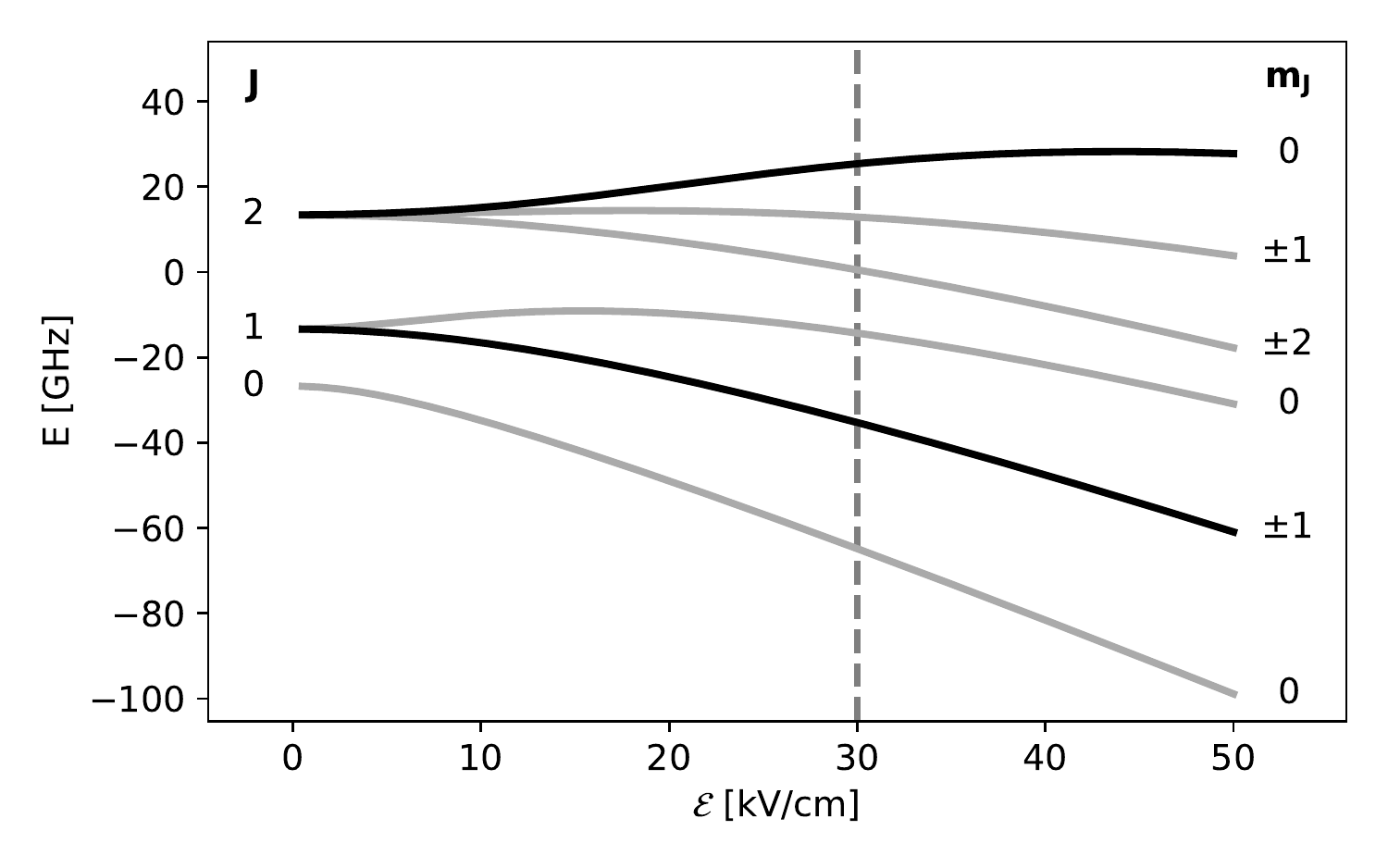}
	\caption{Evolution of the energy eigenstates of the TlF Hamiltonian (Eq. \ref{eq:hyperfine_hamiltonian}) for $\Esca$ ranging from $0\,$V/cm to $50\,$kV/cm, for $J=0,1,2$. States used in \CENTREX\ are shown in bold. Hyperfine structure is unresolved in this plot.}
	\label{fig:low_to_high_field}
\end{figure}

Throughout most of the \CENTREX\ apparatus, TlF molecules experience a non-zero $\Esca$-field and (nominally) zero $\Bsca$-field.
The character of the energy eigenstates changes significantly depending on the $\Esca$-field magnitude, which varies dramatically between stages of the experiment. Hence, it is useful to describe the energy eigenstates of the TlF electronic ground state in different regimes of $\Esca$-field strength (with $\Bsca = 0$), as defined by the ratio of Stark shifts, $\Delta E_{\rm S} = \langle \mathcal{H}_{\rm S}\rangle \sim \mu_e^2 \Esca^2/B$, to the strength of hyperfine interactions, $E_{\rm hf} = \langle \mathcal{H}_{\rm sr} + \mathcal{H}_{\rm ss}\rangle \sim c_j$, or rotational energies, $E_{\rm rot} = \langle \mathcal{H}_{\rm rot}\rangle \sim B$. In all regimes, the total angular momentum projection $m_F$ along 
a space-fixed quantization axis $\hat{z}$ (always defined such that $\Evec$ is very nearly parallel to $\hat{z}$), is an exact quantum number.

In the low-field regime, where $\Delta E_{\rm S} \ll E_{\rm hf}$, energy eigenstates retain $J$, $F$, and $F_1$ as approximate quantum numbers. 
In the mid-field regime, where $E_{\rm hf} \ll \Delta E_{\rm S} \ll E_{\rm rot}$, both $J$ and $m_J$ are approximate quantum numbers. 
Here, the tensor part of the Stark shifts gives rise to energy splittings between levels with different values of $|m_J|$ that are comparable in size to the scalar shifts, i.e., of order $\Delta E_{\rm S}$.  Thus, when $m_J\neq 0$, $\vec{J}$ is strongly coupled to $\Evec$ (and hence to the molecular axis $\vec{\hat{n}}$) by this Stark interaction. In this case, each nuclear spin is coupled to $\vec{J}$ (and thus also to $\Evec$) by the spin-rotation interactions of $\mathcal{H}_{\rm sr}$. Hence, here $m_{I_1}$ and $m_{I_2}$ are approximate quantum numbers. By contrast, in states where $m_J=0$ (including when $J=0$) in this regime, $\left\langle \mathcal{H}_{\rm sr}\right\rangle$ vanishes to first order, and the nuclear spins do not couple to $\vec{J}$ and $\Evec$. However, the nuclear spins remain coupled to each other via the spin-spin interaction $\mathcal{H}_{\rm ss}$.  So, here the total nuclear spin $\vec \It = \vec I_1 + \vec I_2$ and its projection $m_{\It}$ are approximate quantum numbers in addition to $J$ and $m_J=0$. 
Finally, in the high-field regime where $E_{\rm hf} \ll E_{\rm rot} \lesssim \Delta E_{\rm S}$, $J$ states are strongly mixed, and separations between $m_J$ states are on the order of $E_{\rm rot}$. Here, eigenstates are defined by the same approximate quantum numbers as in the mid-field regime, aside from $J$. We refer to these strongly mixed states with the label $\widetilde{J}$, which corresponds to the value of $J$ that any given state connects to adiabatically, if the $\Esca$-field is reduced.
Table \ref{tab:E_field_regimes} summarizes the different regimes and associated eigenstates.  
\begin{table*}[t]
    \centering
    \def\colseplarge{4ex} 
	\begin{tabular}{@{}r@{\hspace{\colseplarge}}r@{$\,\Delta E_{\rm S}\,$}l@{\hspace{\colseplarge}}r@{$\,\Esca\,$}l@{\hspace{\colseplarge}}>{\raggedright\arraybackslash}m{4cm}@{}}
    	\toprule
    	Regime & \multicolumn{2}{c}{Definition} & \multicolumn{2}{c}{Field strength} & Approx. eigenstates \\
		\midrule
		 Low
		 & & $\ll E_{\rm hf}$
		 & & $\lesssim 50 \Vcm$
		 & $\ket{J,F_1,F,m_F}$ \\\addlinespace[2ex]
	     Mid
	     & $E_{\rm hf} \ll$ & $\ll E_{\rm rot}$
	     & $50 \Vcm \ll$ & $\lesssim 5 \kVcm$
	     & $\ket{J,m_J \neq 0}\ket{m_{I_1},m_{I_2}}$ 
	       $\ket{J,m_J=0}\ket{\It,m_{\It}}$ \\\addlinespace[2ex]
	     High
	     & $E_{\rm hf} \ll E_{\rm rot} \lesssim$ &
	     & $5\kVcm \ll$ &
	     & $\ket{\widetilde{J},m_J\neq 0} \ket{m_{I_1},m_{I_2}}$
	       $\ket{\widetilde{J},m_J=0}\ket{\It,m_{\It}}$ \\
		\bottomrule
	\end{tabular}
	\caption{Regimes of electric field strength and associated eigenstates in TlF.}
	\label{tab:E_field_regimes}
\end{table*}

Figure \ref{fig:low_to_mid_field} shows how the relevant energies and eigenstates evolve from the low-field to the mid-field regime for $J=1$ and $J=2$ states.  Bold curves are states directly relevant to \CENTREX. Figure \ref{fig:low_to_high_field} shows a zoom out of states up to $J=2$ from low to high fields.

The $^{205}$Tl NSM measurement is carried out in $\tilde{J} = 1,\, m_J = \pm1$ states of TlF at large electric field $\mathcal{E} = 30$ kV/cm. This choice of states takes advantage of the structure of TlF in electric fields, in two ways. First, the observable energy shift associated with $S$, $\Delta E$, scales linearly with the degree of polarization $\mathcal{P}$ of the TlF molecule (Eq. \ref{eq:frequency_shift_due_to_NSM}). An electric field more easily polarizes states with low $J$, since $\mathcal{P}$ arises from mixing between states with different parity and thus different $J$; these states are closest together when $J$ is small. Additionally, as discussed in Sec. \ref{Sec:InternalComagnetometry}, certain dangerous systematic errors in the NSM measurement are dramatically suppressed in the presence of a strong spin-rotation interaction (later referred to as an effective intra-molecular magnetic field). This requires $m_J \neq 0$. The $\tilde{J} = 1,\, m_J = \pm1$ states hence provide the best combination of sensitivity and systematic error suppression in TlF.\footnote{$|\mathcal{P}|$ is larger in the $J=0,m_J = 0$ states, given the same $\Esca$-field value. Hence, the NSM gives larger energy shifts there. However, in these states where $m_J = 0$, the effective intra-molecular magnetic field vanishes.}

\section{Experiment Overview}
\label{sec:experimental_overview}
\begin{figure*}
	\centering
	\includegraphics[width=\textwidth]{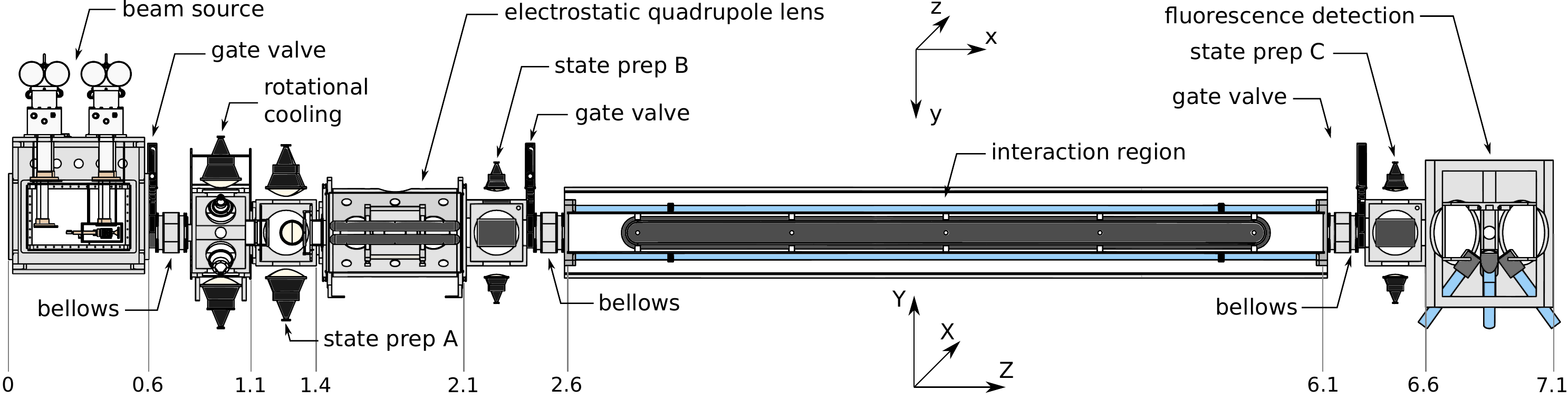}
	\caption{Overview of the planned \CENTREX\ beamline. Distance in meters is shown on the bottom. Modules following the electrostatic quadrupole lens are currently being designed, so few details are given.}
	\label{fig:beamline}
\end{figure*}

\begin{figure*}
	\centering
	\def\svgwidth{\textwidth}
	\includegraphics[width=\textwidth]{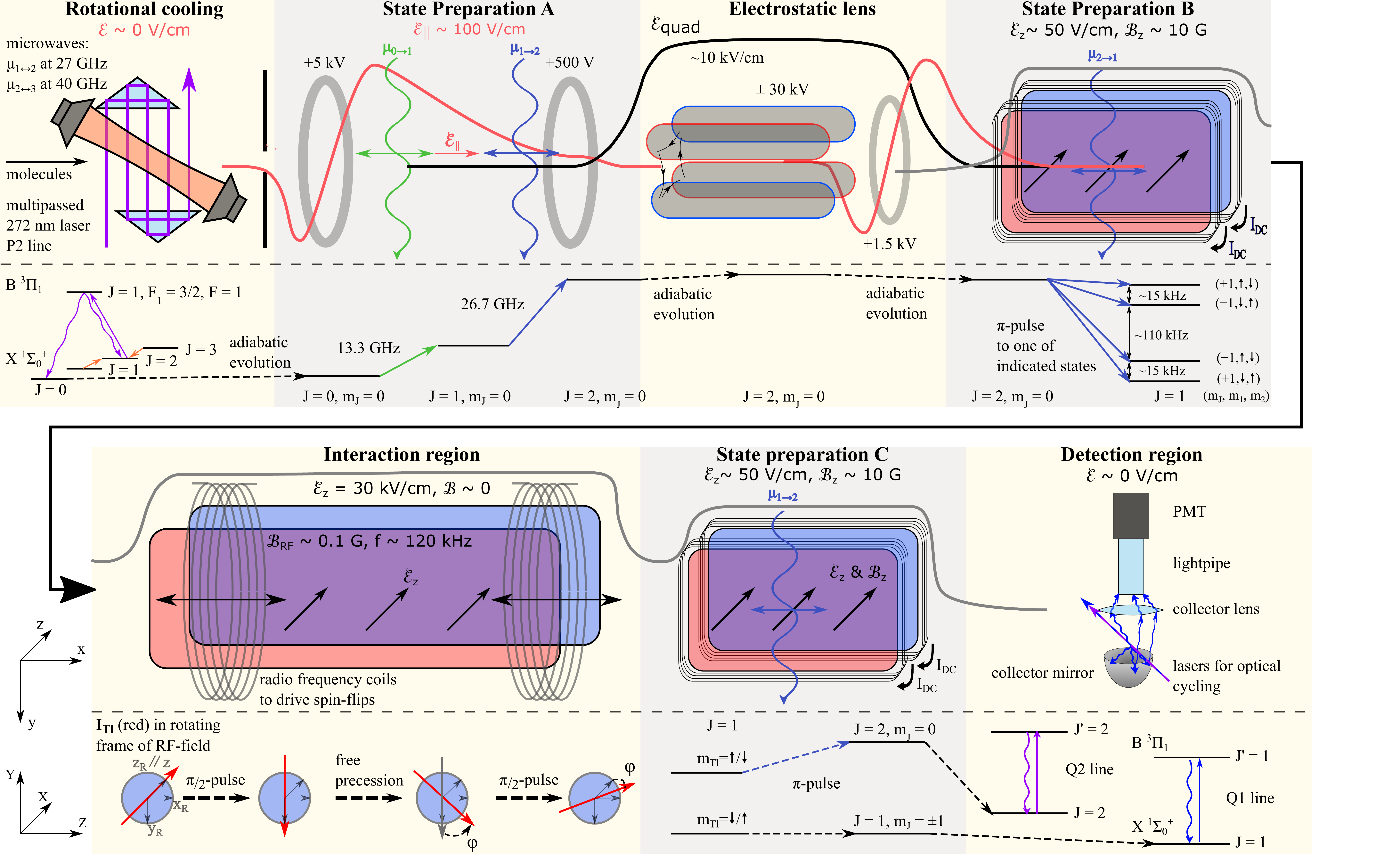}
	\caption{Overview of the regions the TlF molecules traverse as they move through the \CENTREX\ beamline. After emerging from the beam source (not shown), the molecules enter the rotational cooling region. Here, the population in the rotational levels $J = 1$, 2, and 3 is optically pumped to $\ket{J=0,\,F=0}$. (See Sec.~\ref{sec:rotational_cooling} and Fig.~\ref{fig:rotational_cooling_scheme} for more detail.) From there, they move into state preparation region A, where adiabatic passage is applied twice to transfer the molecules to $\ket{J=2,\ m_J=0}$ (Sec.~\ref{sec:state_preparation_region_a}), the state focused by the electrostatic lens (Sec.~\ref{sec:electrostatic_lens}). In state preparation region B (Sec.~\ref{sec:state_preparation_B}), the molecular state is transferred into one of the $\ket{J=1,\ m_J=\pm1}$ states before proceeding to the interaction region (Sec.~\ref{sec:interaction_region}). Here, using NMR with the SOF technique, we transform the frequency shift between Tl spin up and down states, due to $\mathcal{H}_\text{CPV}$ (Eq. \ref{eq:Hamiltonian_effective_interaction}), into a population difference between these states. Subsequently, in state preparation region C (Sec.~\ref{sec:state_preparation_region_c}), one of the Tl spin state populations is transferred to a $J=2$ state. This makes the two populations resolvable with laser-driven transitions, to facilitate state readout (Sec.~\ref{sec:detection_region}). Finally, in the detection region, optical cycling and fluorescence collection are used for efficient, quasi-simultaneous detection of the two populations.  The red, grey, and black curves in the figure indicate the magnitude of the electric field along, respectively, the beam direction $Z$, the interaction region field direction $z$, and the transverse electric quadrupole field directions $X,Y$.}
	\label{fig:beamline_schematic}
\end{figure*}

\CENTREX\ consists of a series of modules. 
In this section we describe each region and its function, following the path of a typical TlF molecule in the experiment. A cryogenic buffer gas \emph{beam source} (BS) forms a cold, slow, and bright molecular beam. Next, a \emph{rotational cooling region} (RC) accumulates molecules from the many thermally populated 
states into a single hyperfine state in $J=0$.
Next, the molecules are coherently transferred from $J=0$ to $J=2,\,m_J=0$ in \emph{state preparation region A} (SPA). This makes it possible to focus the molecular beam at a final detection region downstream with the \emph{electrostatic quadrupole lens} (EQL). \emph{State preparation region B} (SPB) coherently moves population from $J=2,\,m_J=0$ to $J=1,\,m_J=\pm 1$, the states used for the Schiff moment measurement. Next, the \emph{main interaction region} (MI) performs nuclear magnetic resonance (NMR) in the presence of a strong polarizing $\Evec$-field using the technique of separated oscillatory fields (SOF) \cite{PhysRev.78.695,ramsey1951phase}. A short RF magnetic field subregion creates a superposition of thallium nuclear spin states. After a period of free precession in $\Evec$, a second RF field subregion maps the phase accumulation due to the energy difference of the spin states---including a contribution from the Schiff moment---into a population difference between the spin states. In \emph{state preparation region C} (SPC), each spin state is transferred to a different rotational state. Finally, the rotational populations are read out with laser-induced fluorescence (LIF) and optical cycling in the \emph{fluorescence detection region} (FD). An overview of the beamline design is shown in Figures~\ref{fig:beamline} and~\ref{fig:beamline_schematic}.

Because the quantizing fields in \CENTREX~change direction throughout the apparatus, we find it useful to use two different coordinate systems.  We use $(X,Y,Z)$ to denote ``beamline'' coordinates, where $\hat{Z}$ points in the average direction of the molecular beam and $\hat{Y}$ is vertical upward. Similarly, we use $(x,y,z)$ to denote ``interaction region'' coordinates. Here, $\hat{z}$ lies along a line parallel to the average $\Evec$-field in the interaction region, $\langle\Evec\rangle_{\rm MI}$, $\hat{x}$ is the vector closest to the average beam velocity that is also perpendicular to $\hat{z}$, and $\hat{y}$ is the vector closest to downward (along gravity) that is perpendicular to $\hat{z}$ and $\hat{x}$.  
The direction of $\hat{z}$ (parallel or antiparallel to $\langle\Evec\rangle_{\rm MI}$) is fixed in the lab, set by the definitions of $\hat{x}$ and $\hat{y}$ and by demanding a right-handed coordinate system.  Hence, the $\Evec$ field in the MI region is (nominally) $\Evec = \Esca \hat{z}$, where $\Esca$ can take either sign.

\subsection{Beam Source}
\label{sec:beamsource}
The source of the TlF molecules is a cryogenic neon buffer gas beam \cite{hutzler2012buffer}. A copper cell containing a solid TlF target is cooled to $18\,$K. The TlF is ablated with a pulsed Nd:YAG laser operating at up to $50\Hz$, while Ne flows continuously through the cell at a typical rate of $40\sccm$. A $4\kelvin$ layer surrounds the cell and cryopumps the Ne.  Ablated TlF reaches thermal equilibrium with the Ne buffer gas before the cell exit, where the beam cools further as it expands into vacuum. The cold cell aperture ($6.35\mm$ diameter) defines the zero position along the beamline axis, $\vec{Z}$. Two $25.4\mm$ diameter apertures (one in the $4\kelvin$ layer at $Z=43\mm$, one in a blackbody shield at $Z=81\mm$) collimate the beam. 

The velocity distributions of the TlF beam were measured as follows. An additional collimator 
was placed downstream. After this collimator, a laser beam, tuned to a $Q1$ line of the $X-B$ transition, crossed the molecular beam. Here, laser-induced fluorescence (LIF) was recorded with a photomultiplier tube (PMT). The LIF signal as a function of laser detuning, with laser beams perpendicular to or at $45^\circ$ to the TlF beam, yielded information on the velocity distributions. The longitudinal distribution is very nearly Gaussian, with mean $\langle v_Z \rangle = 184(17)$ m/s and Gaussian width $\sigma_{v_Z} = 16.1(8)$ m/s. The latter corresponds to translational temperature $T_\text{tr}=7.0(7)$ K.
The TlF beam divergence was determined from the shape of an isolated $Q$-branch ($\widetilde{J}'= J$) absorption line,  probed upstream of any collimation. The FWHM spread in transverse velocity here was $93(3)$ m/s, corresponding to a divergence cone half-angle of $14.0(1.5)^\circ$.

\begin{figure}
	\centering
	\includegraphics[width=0.48\textwidth,unit=1mm]{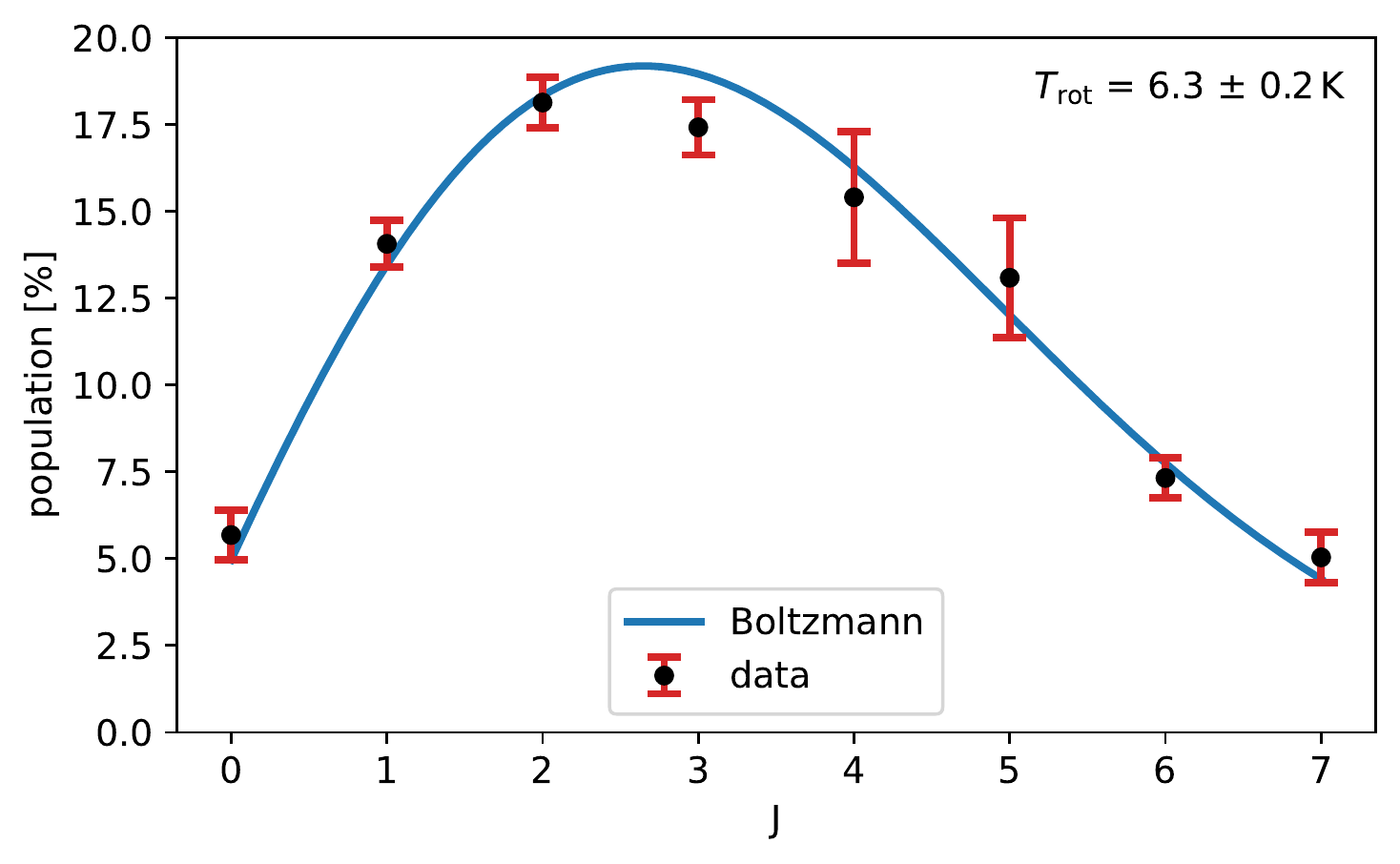}
	\caption{Relative rotational state populations of TlF in the \CENTREX\ beam source with typical conditions as described in the text, overlaid with a fit to a Boltzmann distribution. From the fit, we find the rotational temperature $T_\mathrm{rot} = 6.3(2)$\,K.}
	\label{fig:rotational_temperature}
\end{figure}
The rotational temperature was determined by measuring the population in different rotational states via the size of LIF signals on $R$-branch transitions, where $\widetilde{J}' = J+1$.  The laser can resolve hyperfine structure in the excited but not in the ground state. Targeting the excited-state sublevel with the largest possible angular momentum, $\widetilde{F}' = \widetilde{J}'+ 1 = J+2$, ensures that only a single ground state hyperfine sublevel, with $F=J+1$, is excited. This considerably simplifies the extraction of rotational level populations from LIF signals. Subsequently, the relative populations are fit to a Boltzmann distribution,
\begin{equation}
    P(J)= g(J)\exp\left(-\frac{BJ(J+1)}{\kB T_\text{rot}}\right),
\end{equation}
where $g(J) = 4(2J+1)$ is the degeneracy of each rotational level. This procedure, illustrated in Figure \ref{fig:rotational_temperature}, yielded $T_\mathrm{rot} = 6.3(2)$\,K. 

From known line strengths~\cite{norrgard2017hyperfine,PhysRevA.101.042506,clayburn2020measurement}, calculated solid angle of fluorescence detection, and calibrated PMT sensitivity, we found a time-averaged TlF beam intensity of $5\times 10^{12}$ molecules/state/sr/s. Here, each $m_F$ sublevel is considered one state, and the time average is taken over 1 second when operating at 50 Hz pulse repetition rate. This is comparable to intensities found in other cryogenic buffer gas beam sources \cite{hutzler2012buffer}.

\subsection{Rotational Cooling}
\label{sec:rotational_cooling}
\begin{figure*}
	\centering
	\small
	\begin{minipage}[c]{0.49\textwidth}
		\begin{tikzpicture}

    \node (J0) at (0,0) {$\ket{J=0^+}$};
    \node (J1) at (2,.5) {$\ket{J=1^-}$};
    \node (J2) at (4,1) {$\ket{J=2^+}$};
    \node (J3) at (6,1.5) {$\ket{J=3^-}$};
    \node (Je1) at (3,4) {$\ket{\widetilde{J}=1^-,\widetilde{F_1}=3/2,\widetilde{F}=1}$};

    \node[below=-5pt of J0] (p0) {\footnotesize5.0~\%};
    \node[below=-5pt of J1] (p1) {\footnotesize13.4~\%};
    \node[below=-5pt of J2] (p2) {\footnotesize18.3~\%};
    \node[below=-5pt of J3] (p3) {\footnotesize19.0~\%};

    \node (GS) at (7.2,.75) {$X^1\Sigma^+$};
    \node (ES) at (7.2,4) {$B^3\Pi_1$};

    \draw[-{Latex[length=4mm,width=4mm]},line width=3pt,gray] (4,1.25) -- node[above,black,sloped]{$P2$ $F1$} (3.25,3.75);

    \tikzset{snake it/.style={-{Latex[length=2mm,width=2mm]}, decorate, decoration={snake,amplitude=2pt,pre length=2pt,post length=3pt}}}
    \draw[snake it] (2,3.75) -- node[left,xshift=-1pt]{\footnotesize0.484} (0,.25);
    \node[above=3 of J0] {\normalsize \textbf{a)}};
    \draw[snake it] (2.75,3.75) -- node[left,xshift=3pt,yshift=-10pt]{\footnotesize0.516} (3.5,1.25);

    \tikzset{biarrow/.style={{Latex[length=1.5mm,width=1.5mm]}-{Latex[length=1.5mm,width=1.5mm]}}}
    \draw[biarrow] (p1) to[bend right] node[below,sloped]{\small26.7~GHz} (p2);
    \draw[biarrow] (p2) to[bend right] node[below,sloped]{\small40.0~GHz} (p3);

\end{tikzpicture}
	\end{minipage}
	\hfill
	\begin{minipage}[c]{0.49\textwidth}
		\begin{tikzpicture}[scale=4]
    \def\len{.05}

    \def\pos{-18}
    \def\lenmark{.025}
    \def\xoffset{.18}

    \def\xa{-4.5*\len}
    \def\xb{-2.5*\len}
    \def\xc{-0.5*\len}
    \def\xd{+1.5*\len}
    \def\xe{+3.5*\len}
    \def\xf{-6.5*\len}
    \def\xg{+5.5*\len}

    \def\xexc{.8}
    \def\yexc{.6}
    \draw (\xexc+\xc, \yexc-0.003325) node (exc) {} -- (\xexc+\xc+\len, \yexc-0.003325);
    \draw (\xexc+\xb, \yexc-0.003325) -- (\xexc+\xb+\len, \yexc-0.003325);
    \draw (\xexc+\xd, \yexc-0.003325) -- (\xexc+\xd+\len, \yexc-0.003325) node (exc2) {};
    \node[align=center] (Fp1) at (\xexc+\xc, \yexc-0.003325+0.1) {$\ket{\widetilde{J}=1^-,\widetilde{F_1}=3/2,\widetilde{F}=1}$};

    \def\yB{.2}
    \draw (\xc, \yB-.217455) -- (\xc+\len, \yB-.217455);
    \draw (\xb, \yB-.217455) -- (\xb+\len, \yB-.217455);
    \draw (\xd, \yB-.217455) -- (\xd+\len, \yB-.217455);
    \draw (\xc, \yB-.172936) -- (\xc+\len, \yB-.172936);
    \draw (\xb, \yB-.172936) -- (\xb+\len, \yB-.172936);
    \draw (\xd, \yB-.172936) -- (\xd+\len, \yB-.172936);
    \draw (\xa, \yB-.172936) -- (\xa+\len, \yB-.172936);
    \draw (\xe, \yB-.172936) -- (\xe+\len, \yB-.172936);
    \draw (\xc, \yB+.105876) -- (\xc+\len, \yB+.105876);
    \draw (\xb, \yB+.105876) -- (\xb+\len, \yB+.105876);
    \draw (\xd, \yB+.105876) -- (\xd+\len, \yB+.105876);
    \draw (\xa, \yB+.105876) -- (\xa+\len, \yB+.105876);
    \draw (\xe, \yB+.105876) -- (\xe+\len, \yB+.105876);
    \draw (\xc, \yB+.141095) -- (\xc+\len, \yB+.141095);
    \draw (\xb, \yB+.141095) -- (\xb+\len, \yB+.141095);
    \draw (\xd, \yB+.141095) -- (\xd+\len, \yB+.141095);
    \draw (\xa, \yB+.141095) -- (\xa+\len, \yB+.141095);
    \draw (\xe, \yB+.141095) -- (\xe+\len, \yB+.141095);
    \draw (\xf, \yB+.141095) -- (\xf+\len, \yB+.141095);
    \draw (\xg, \yB+.141095) -- (\xg+\len, \yB+.141095);
    \node[align=center] (J2) at (\xd-0.5*\len,\yB+.141095+.1) {$\ket{J=2^+}$};
    \node[above=0.25 of J2] {\normalsize \textbf{b)}};

    \def\xgnda{.8}
    \def\xgndb{1.3}
    \draw (\xgnda+\xc, -0.003325) -- node (gnd1) {} (\xgnda+\xc+\len, -0.003325);
    \draw (\xgnda+\xb, -0.003325) -- (\xgnda+\xb+\len, -0.003325);
    \draw (\xgnda+\xd, -0.003325) -- (\xgnda+\xd+\len, -0.003325);
    \draw (\xgndb+\xc, 0.009975)  -- node (gnd0) {} (\xgndb+\xc+\len, 0.009975);

    \def\xF{7.5*\len}
    \def\yF{.01}
    \node (F3)  at (\xF,\yB+.141095+\yF) {3};
    \node (F2a) at (\xF,\yB+.105876-\yF) {2};
    \node (F2b) at (\xF,\yB-.172936+\yF) {2};
    \node (F1)  at (\xF,\yB-.217455-\yF) {1};
    \node[above=-0.03 of F3]   {$F$};
    \def\xFf{0.10}
    \node[below=\xFf of gnd0] {$F=0$};
    \node[below=\xFf of gnd1] {$F=1$};
    \node[align=center] (J0) at (0.5*\xgnda+0.5*\xgndb,-2*\xFf) {$\ket{J=0^+}$};

   \draw [decorate,decoration={brace,amplitude=5pt,mirror,raise=5pt}] (\xF,\yB-.217455-2*\yF) -- (\xF,\yB+.105876) node [black,midway,anchor=west,xshift=5] (trans) {};
   \draw[-{Latex[length=3mm,width=3mm]},line width=2pt,gray] (trans) -- node[above,black,sloped]{$P2~F1$} (exc);
   \tikzset{snake it/.style={-{Latex[length=2mm,width=2mm]}, decorate, decoration={snake,amplitude=1.5pt,pre length=1pt,post length=2pt}}}
   \draw[snake it] (exc2) -- node[right,xshift=0pt]{0.337} (gnd0);
   \draw[snake it] (exc2) -- node[left,xshift=0pt] {0.147} (gnd1);
\end{tikzpicture}
	\end{minipage}
	\mbox{}
	\caption{Rotational cooling scheme. \part{a} The thick solid arrow marks a UV laser driving the $P2$ $F1$ transition; bent arrows represent microwaves, and wavy arrows indicate spontaneous emission.  Rotational branching ratios are calculated with data from \cite{norrgard2017hyperfine}. The odd-parity $\tilde{J}' = 1^-$ excited state can only decay to states with $J=0^+,2^+$. Percentages under the ground-state kets are the thermal population at temperature $T_{\rm rot}=6.3\,$K, prior to rotational cooling. \part{b} Hyperfine structure relevant to optical pumping. Decays back to $J=2^+$ are not shown. The $P2$ $F1$ transition does not excite $\ket{J=2^+,F=3}$. The nearest optical transition that couples to the $J=2^+$ hyperfine manifold is separated from this line by about 550 MHz.}
	\label{fig:rotational_cooling_scheme}
\end{figure*}
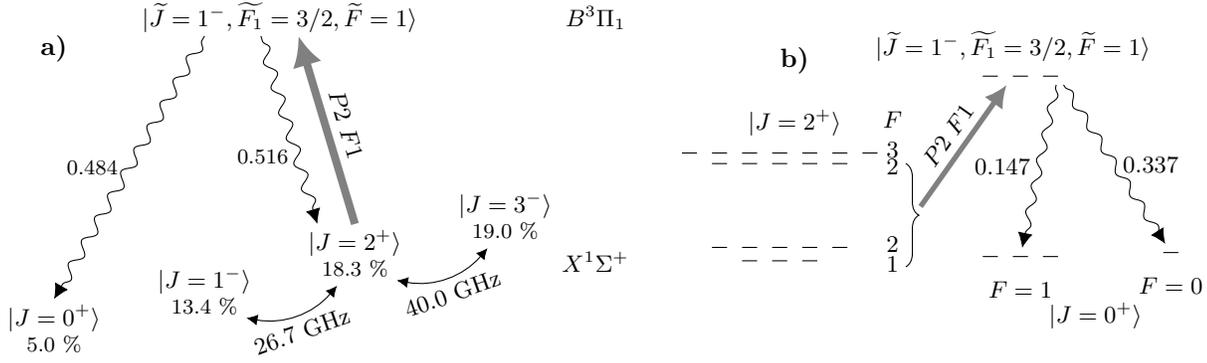
In a Boltzmann distribution at $T_\text{rot} = 6.3$ K, about 50\% of the TlF population is in states with $J=0$ through $J=3$. To maximize the Schiff moment measurement sensitivity in \CENTREX\, this population is dissipatively pumped to the $J=0, F=0$ level, which becomes the initial state for all further steps in the experiment. 
This rotational cooling will be accomplished using a single optical pumping laser and two microwave driving fields. The laser couples the $J=2$ state to an excited state with $\widetilde{J}'=1$. We calculate that about half of the decays from the excited $\widetilde{J}'=1$ state end in the $J=0$ state; nearly all of the remainder returns to the $J=2$ state. Branching to other vibrational states is $\lesssim\! 1\%$ \cite{norrgard2017hyperfine, hunter2012prospects}.
The microwaves resonantly couple $J=1\!\leftrightarrow\! 2$ and $J=2\! \leftrightarrow\! 3$. Repeated excitation-decay cycles then lead to accumulation of population from $J=1,~2,\mathrm{and}~3$ into the $J=0$ state, as shown in Fig.~\ref{fig:rotational_cooling_scheme}. 

The presence of hyperfine structure adds considerable complexity to rotational cooling in TlF. The aim is to accumulate population in the $J=0,~F=0$ hyperfine level. While the ground-state hyperfine splitting is smaller than the laser linewidth, the excited-state hyperfine levels are well separated. We tune the $J=2$ optical pumping laser to resonance with the   $\widetilde{J}'=1,~\widetilde{F_1}'=3/2,~\widetilde{F}'=1$ level. We refer to this line as the $P2~F1$ transition. (As usual, this $P$-branch transition has $\widetilde{J}^\prime = J-1$.) 
Without a considerable effort, this level structure will only support an extremely low excitation and pumping rate due to the formation of long-lived coherent dark states \cite{BerkelandBoshier2002} within the manifold of unresolved ground-state hyperfine and Zeeman sublevels. In \CENTREX, these dark states will be rapidly destabilized \cite{BerkelandBoshier2002} by switching the polarizations of the laser and both microwave beams \cite{ShumanRadiative,YeMicrowaveRemix}, and ensuring that no pair of the three beams are either parallel or perpendicular to each other.\footnote{Due to the selection rule $\left(\Delta F = 0,\pm1\right)$, the $\left(J=2,F=3\right)$ state is dark with respect to laser excitation in this scheme. However, even the population in this state can eventually be optically pumped, since the strong $J=2\!\leftrightarrow\!J=3$ microwave field couples this state to the $J=2,F=2,2,1$ states via Raman-type transitions.}
Even with these measures, the excitation rate on the laser-driven transition is bounded by $\Gamma_{\rm sc} \lesssim \gamma_B\cdot n_\text{e}/(n_\text{g}+n_\text{e})$ \cite{Tarbutt_2013}, where $n_\text{e}=3$ is the number of excited state sublevels, $n_\text{g} = 60$ is the number of simultaneously coupled ground-state sublevels, and $\gamma_B \approx \times 1.6 $ MHz.

We have performed numerical simulations to estimate the efficiency of this rotational cooling scheme, by solving the optical Bloch equations for the full 67-level system coupled by microwaves, lasers, and spontaneous emission.  Our simulations include a realistic transverse velocity distribution; from trajectory simulations, we find that the Doppler shifts experienced by molecules that can enter the lens are as large as $\pm 14\,\gamma_B$. This range is difficult to saturate with simple power broadening, given the limited laser power available in the UV.  To use this power most effectively, we add many sidebands to broaden the laser frequency such that it roughly matches the full Doppler width. In the simulations, we phase-modulate the laser (with its carrier frequency set to resonance with zero-velocity molecules) at frequency $f_{\textrm{mod}}=\gamma_B = 1.6\,$MHz and with modulation depth $\beta_{\textrm{mod}}=8.5$.  (These parameters can be reached with a commercial electro-optic modulator.) We assume a realistic laser intensity ($\sim\!6$ mW/mm$^2$) in a beam multi-passed across the molecular beam to achieve total interaction length of $\sim\! 1 \cm$ (corresponding to 50~$\mu$s interaction time), and 100 mW of microwave power in $2.5 \cm$ diam.\ beams at each frequency. 

Under these conditions, we find that $78\%$ of all molecules originally in the $J=\hbox{0--3}$ states are accumulated into sublevels of the $J=0$ state. Of these, $\sim 50\%$ are in the desired $\ket{J=0,~F_1=\frac{1}{2},~F=0,~M_F=0}$ hyperfine state. This corresponds to a 24-fold increase compared to the initial thermal population in this state, due to rotational cooling. Based on results from simulations under other conditions, and recent demonstrations of extended multipass geometries \cite{privateHunter2019}, we believe that nearly complete pumping can be achieved by extending the interaction length to $5 \cm$. This could increase the desired state population by another factor of 1.28, and also dramatically reduce residual populations in excited rotational levels (which can contribute to background signals and/or systematic errors). Both experimental and further numerical tests of the rotational cooling are ongoing.

\subsection{State Preparation Region A}
\label{sec:state_preparation_region_a}

\begin{figure}
    \centering
    \includegraphics[width = \textwidth/2]{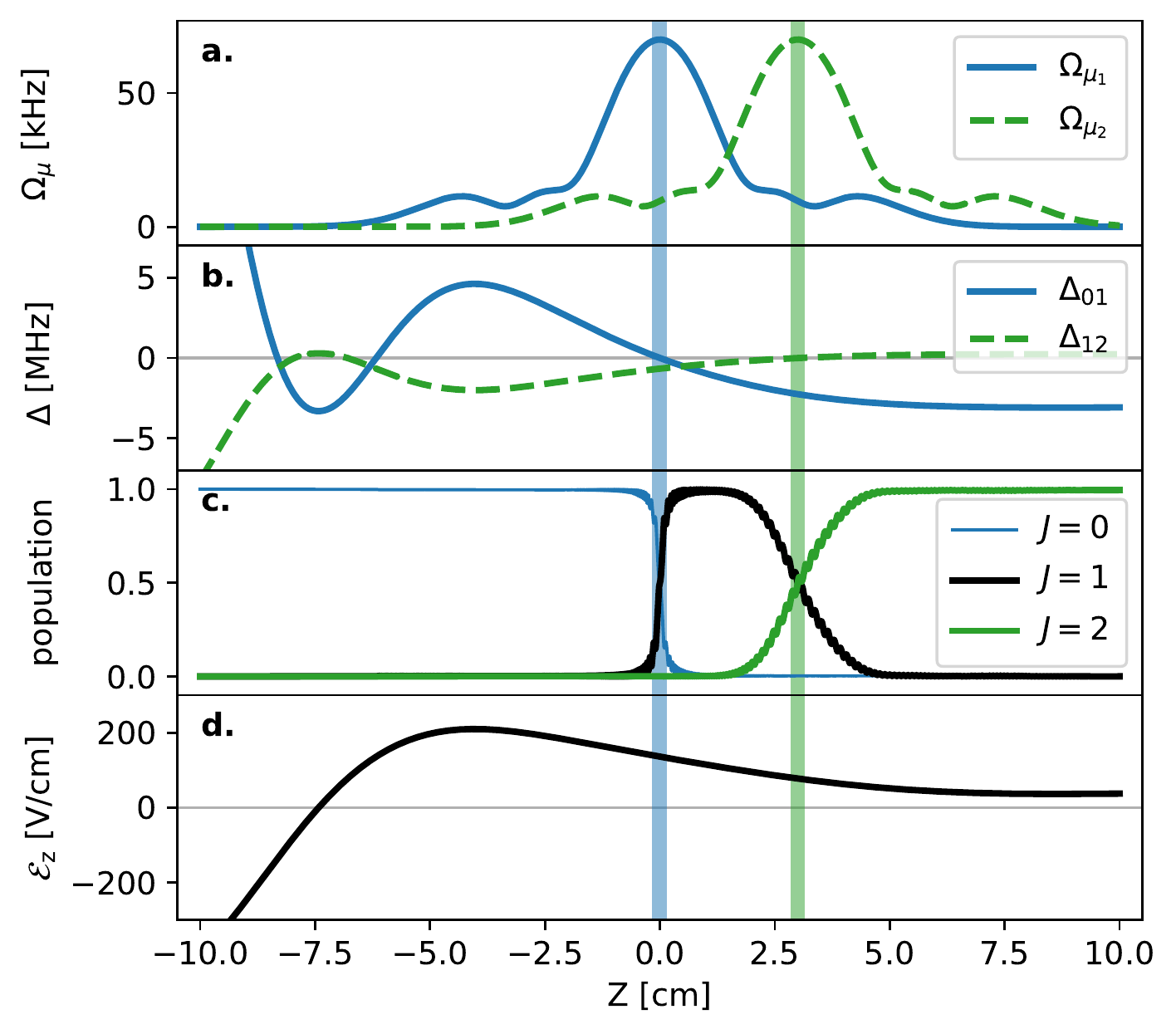}
    \caption{Rabi rates, detunings, state populations and field amplitudes versus position in state preparation region A, where $\hat{Z}$ is the molecular beam direction. \part{a} Calculated Rabi rates $\Omega(Z)$, based on the measured intensity profile from the spot-focusing horns. \part{b} Stark-shifted detunings $\Delta_{01}$ and $\Delta_{12}$ of the transitions $J=0\!\leftrightarrow\! 1$ and $J=1\!\leftrightarrow\! 2$,  respectively. \part{c} Calculated populations of relevant states as the molecules travel through the SPA region, showing a simulated transfer efficiency from $J=0$ to $J=2$ of 99\%. \part{d} Electric field $\Esca_\mathrm{Z}$, based on finite element simulations.}
    \label{fig:state_prep_region_a}
\end{figure}
The electrostatic quadrupole lens in \CENTREX\ is designed to focus molecules in the $\ket{J=2,~m_J = 0}$ state (see Sec.~\ref{sec:electrostatic_lens}). After rotational cooling, a majority of the population is in the state $\ket{J=0,F=0}$, which is a pure $\It=0$ (singlet) state of the nuclear spins. In state preparation region A, the population of this state is coherently transferred to a $\ket{J=2,m_J=0}$ state using a two-stage adiabatic passage (AP) protocol. In each stage, $J$ increases by one and $m_J$ is unchanged, while (nominally) $\It=0$ throughout.

The driving field is provided by two CW, single\hyph frequency free-space microwave beams, tuned to near resonance with the $J=0\rightarrow J=1$ and $J=1\rightarrow J=2$ transitions. The beams are produced by spot-focusing horns, spatially offset so the beam profiles have no significant overlap. The time-varying detuning of each beam from its respective resonance is provided by the quadratic Stark shift due to a spatially varying DC electric field as the molecules fly through the region. The desired $\Delta m_J=\Delta\It=0$ transitions are selectively driven by $\pi$-polarized microwaves. Due to geometric constraints, this requires the DC electric field, $\Evec$, to lie along the molecular beamline, $\hat{Z}$. Fig.~\ref{fig:beamline_schematic} has a schematic overview of this region.

For efficient population transfer via AP, the adiabaticity condition must be fulfilled \cite{budker2004atomic}:
\begin{equation}
	\frac{d\Delta}{dt} \ll \Delta^2 + \Omega_\text{$\mu$w}^2,
\end{equation}
where $\Delta$ is the detuning and $\Omega_\text{$\mu$w}$ the Rabi rate of the microwave drive. Furthermore, the detuning at large times before and after the AP interaction must be larger than the microwave Rabi rate. This is accomplished with fields as shown in Fig.~\ref{fig:state_prep_region_a}.
We simulated the TlF state evolution in the SPA region with peak Rabi rates $\Omega_\text{$\mu$w}=70\kHz$, perfectly pure $\pi$-polarization, microwave intensity profile as measured from the focusing horns, $\Esca_Z$ field from finite element calculations, and including the effect of the Earth's magnetic field. With these assumptions, we found a state transfer efficiency of 99\%.

We are confident that high transfer efficiency can also be reached in the real experiment. Adiabatic passage is a threshold process in that as long as the adiabaticity condition is fulfilled, the state transfer occurs with an efficiency close to 100\% \cite{budker2004atomic}. 
By making sure that the adiabaticity condition is satisfied with a sufficiently safe margin, the effects of various factors that might be expected to lower the efficiency can be mitigated. We have simulated the effect of numerous realistic imperfections such as spatial inhomogeneity of the microwaves and the $\Esca$-field, Doppler shifts, and microwave polarization misalignment; we find that with the available microwave intensity, the efficiency is not noticeably degraded in simulations.

\subsection{Electrostatic Quadrupole Lens}
\label{sec:electrostatic_lens}

\begin{figure}
	\centering
	\begin{overpic}[width=\textwidth/2,unit=1mm]{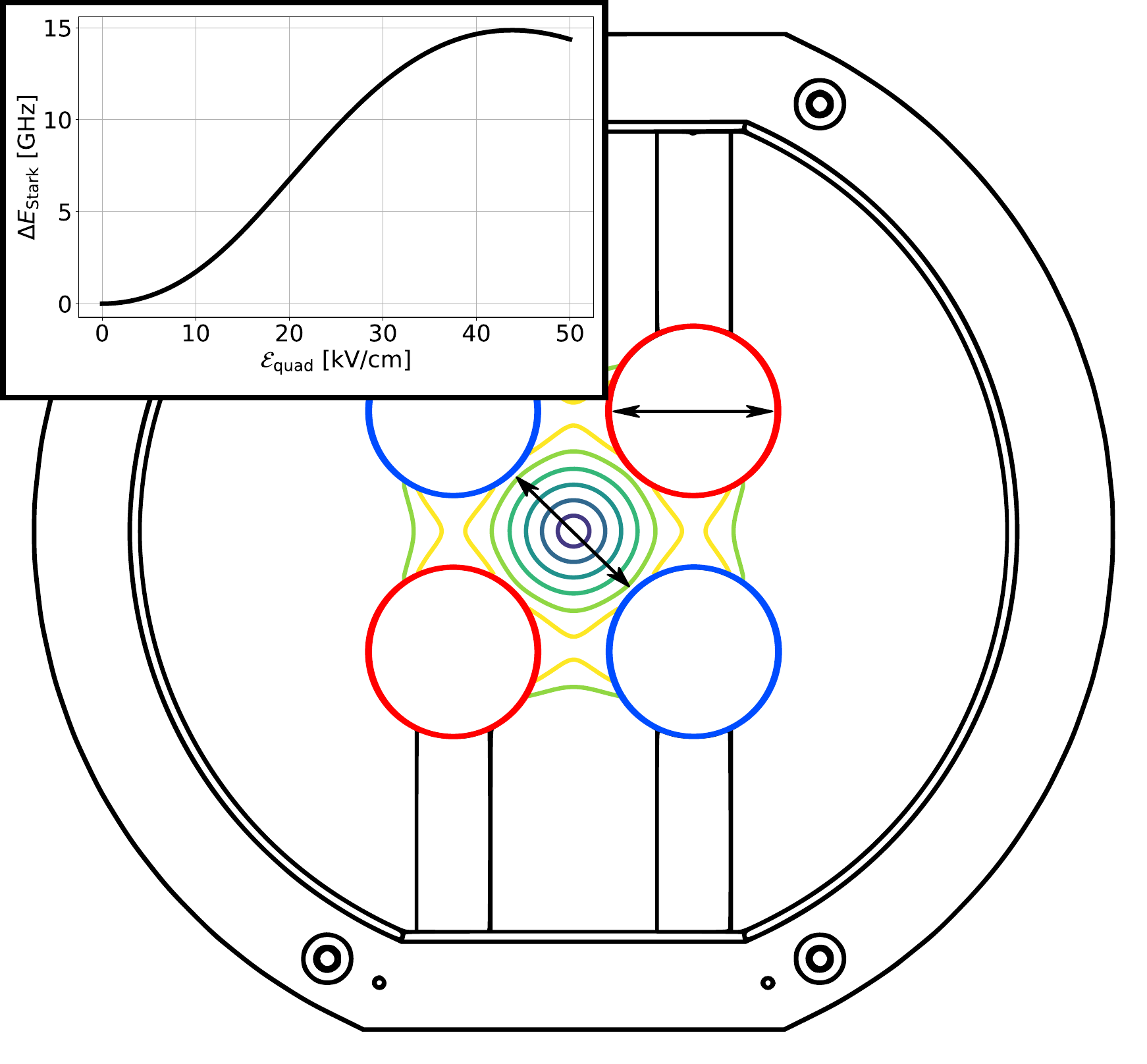}
	    \put(47.5,40.5) {\small \colorbox{white}{$2R=1.75$"}}
		\put(52.5,52) {\small $2R$}
		\put(52,47.5) {\small $-V$}
		\put(52,31) {\small $+V$}
		\put(33,31) {\small $-V$}
		\put(33,49) {\small $+V$}
		\put(1,54) {\large \textbf{a)}}
		\put(6,6) {\large \textbf{b)}}
	\end{overpic}
	\caption{\textbf{a)} Stark shift of the $J=2,\ m_J=0$ hyperfine manifold of states. 
	\textbf{b)} Front view of the electrostatic quadrupole lens. Colored curves are equipotential surfaces. The electrodes have length $l=60\cm$, and applied potentials $\pm V$ up to $\pm30\,$kV. The electrode support structure is 
	mounted on translation stages (not shown) that allow for alignment of the lens under vacuum.}
	\label{fig:quadrupole_lens}
\end{figure}
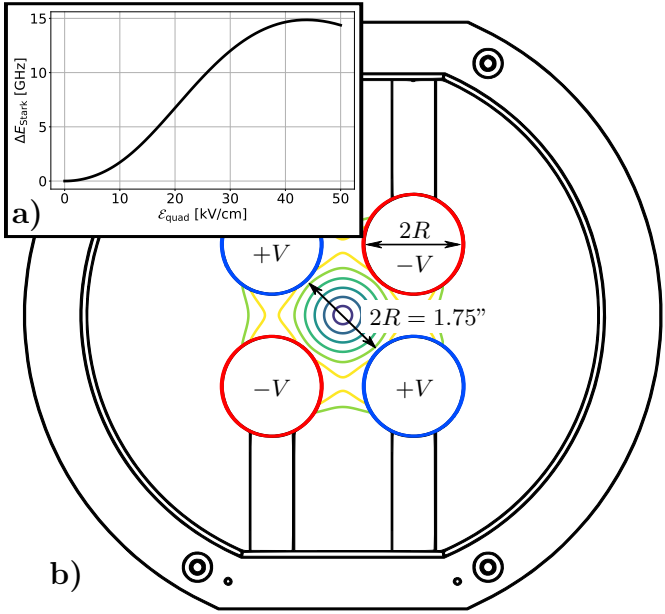
The molecular beam exiting the source is spread over a wide solid angle, so the beam intensity decreases as the square of the distance from the source. The total distance from beam source to final detection in \CENTREX\ is $\approx6.4\,$m, so beam focusing can substantially improve the signal strength. To accomplish this, an electrostatic quadrupole lens (EQL) will be employed.

An EQL with four equidistant cylindrical electrode rods, held at alternating positive and negative potentials of the same magnitude $V$, generates an electric quadrupole field of magnitude 
\begin{equation}
    \left|\mathcal{E}_\mathrm{quad}(r)\right| = \frac{2Vr}{R^2},
\end{equation} 
where $2R$ is both the bore diameter of the lens and the electrode diameter, and $r$ is the distance from the central axis of the quadrupole. A front view of the lens is shown in Fig.~\ref{fig:quadrupole_lens}b. The $J=2,~m_J=0$ states in TlF have a quadratic Stark shift in fields up to $\Esca \approx 20\kVcm$, and slightly slower than quadratic to $\sim 30\kVcm$ (Fig.~\ref{fig:quadrupole_lens}a). For electrode potentials of $\pm V = \pm 30\kV$, the fields inside the bore of the lens do not exceed $30\kVcm$; hence, most molecules in the lens bore remain in the quadratic Stark shift regime. 

A quadrupole field acting on molecules with a quadratic Stark shift produces a harmonic potential along the radial direction within the lens. Under these conditions, the electrostatic lens acts as an analogue to a thick optical lens \cite{bennewitz1955fokussierung}: the lens can be thought of as imaging the molecular beam from the source to the detection region. The trajectories of the molecules can be described by the ray transfer matrices of a thick optical lens \cite{berg1965determination, cho1991tight}:
\begin{equation}
    \begin{gathered}
    \begin{pmatrix}
		r_i \\
		\dot{r_i}/v_Z
	\end{pmatrix}
	= \\
	\begin{pmatrix}
	    1 & z'_i \\
	    0 & 1 
	\end{pmatrix}
	\begin{pmatrix}
		1 & 0 \\
		-f^{-1} & 1
	\end{pmatrix}
	\begin{pmatrix}
	    1 & z'_o \\
	    0 & 1 
	\end{pmatrix}
	\begin{pmatrix}
		r_o\\
		\dot{r_o}/v_Z
	\end{pmatrix},
	\end{gathered}
\end{equation}
where $r_{o/i}$ is the radial position of the molecule in the object/image plane (in our case beam source/detection region), $z'_{o,i}$ is the distance from the object/image plane to the entrance/exit principal plane of the lens, $v_Z$ is the molecular velocity component along the beam direction (i.e.\ longitudinal velocity), and $f$ is the effective focal length given by
\begin{equation}
    \label{eq:EQL_focal_length}
	f = \frac{1}{p\sin\left( pl\right)},~p=\left(\frac{8CV^2}{R^4 m v_Z^2} \right)^{1/2}.
\end{equation}
Here, $l$ is the length of the quadrupole lens, $m$ the molecular mass, and $C$ is a constant corresponding to the strength of the quadratic Stark shift for a given $J,~m_J$ state \cite{brown2003rotational}:
\begin{equation}
    C = \frac{\mu_e^2}{2hB}\frac{J\left(J+1\right)-3m_J^2}{J\left(J+1\right)\left(2J-1\right)\left(2J+3\right)},
\end{equation}
where $B$ is the rotational constant (Tab.~\ref{tab:hyperfine_hamiltonian}) and $\mu_e$ the molecular electric dipole moment. As shown in Eq. \ref{eq:EQL_focal_length}, the focal length depends on the velocity of the molecules. The spread of longitudinal velocities in the molecular pulse thus gives a range of focal lengths; this chromatic aberration increases the focal spot size. Aberrations due to deviation of the Stark shift from a purely quadratic spatial dependence have a similar effect.

Due to the complexity added by the aberrations, the length and bore diameter of the lens were optimized with Monte Carlo simulations of molecular trajectories through the entire apparatus. These simulations were done before much of the beamline was designed and prior to measurements of the molecular beam properties, and thus educated guesses had to be made for the parameters. For the molecular beam we assumed a Gaussian longitudinal velocity distribution with $\langle v_Z \rangle = 200$ m/s and $\sigma_{v_Z} = 13$ m/s. The beamline was taken to have a distance of 0.81 m from the molecular source to the start of the lens, and 3.63 m from the end of the lens to detection. The detection region was taken to have an acceptance area of $10\mm\times30\mm$. The source was taken to have a diameter of $20\mm$, and was located $0.25"$ downstream from the cold cell exit aperture. This was based on an estimate of the molecular cloud size at the 'zone of freezing' where interactions between molecules are assumed to have ceased \cite{hutzler2012buffer}. The length and diameter of the lens were then optimized by maximizing the expected number of detected molecules when the electrodes were at $\pm 30\kV$. The optimal combination was found to be a diameter $2R= 1.75$", and a length $l=60\cm$. The simulated gain in the number of molecules making it to detection was a factor of 24.

Some of the beamline and molecular beam properties are now known better than when the lens was designed. The measured molecular beam velocity is slightly lower at $\langle v_z \rangle = 184\,$m/s (Sec.~\ref{sec:beamsource}) than previously assumed. To compensate for the lower velocity, the electrode voltages are lowered to $\pm27$ kV. The source-to-lens-distance is 1.01 m, and the lens-to-detection-distance 4.45 m (liable to change by $\sim 10$ cm as the SPB and SPC regions are designed).  With these parameters, the simulated gain in the number of molecules reaching the detection region is a factor of $23.2 \pm 0.9$ where the uncertainty is based on Poisson statistics in the simulation.

\subsection{State Preparation Region B}
\label{sec:state_preparation_B}
After the electrostatic lens, TlF resides in a $J=2,~m_J=0$ state, but the NSM measurement requires molecules to be in a $J=1$ state with $m_J=\pm1$ \cite{wilkening1984search,hinds1980experiment} (see Sec. \ref{sec:TlF_in_E_fields}). The required state transfer takes place in state preparation region B. To achieve this, a resonant microwave field with $x$-polarization (in ``interaction region'' coordinates) will be applied in the presence of a magnetic field, $\Bsca_{\rm SPB} \approx 10$ G, and a quantizing electric field, $\Esca_{\rm SPB} \approx 50~\Vcm$. The $\Bvec_{\rm SPB}$-field is parallel to $\Evec_{\rm SPB} = \Esca_{\rm SPB}\hat{z}$, and acts to distinguish $\pm m_J$ states. Here, adiabatic passage can drive undesired transitions to unwanted states nearby in energy. So, here we use a microwave $\pi$-pulse to achieve optimized transfer efficiency. With a peak Rabi rate $\Omega_\text{$\mu$w} = 1.5\kHz$, the simulated transfer efficiency is $\sim\! 96\%$. In practice, the transfer efficiency is likely to be reduced due to non-uniformities of the electric and magnetic fields, since changes in the fields cause the transition frequencies to shift away from the microwave frequency. To achieve the quoted 96\% efficiency, $\Bsca$ needs to be uniform to within $\delta\Bsca/\Bsca < 10^{-3}$ and $\Esca$ to within $\delta\Esca/\Esca < 10^{-4}$. We are designing coils and electrodes to meet these specifications.

\subsection{Main Interaction Region}
\label{sec:interaction_region}

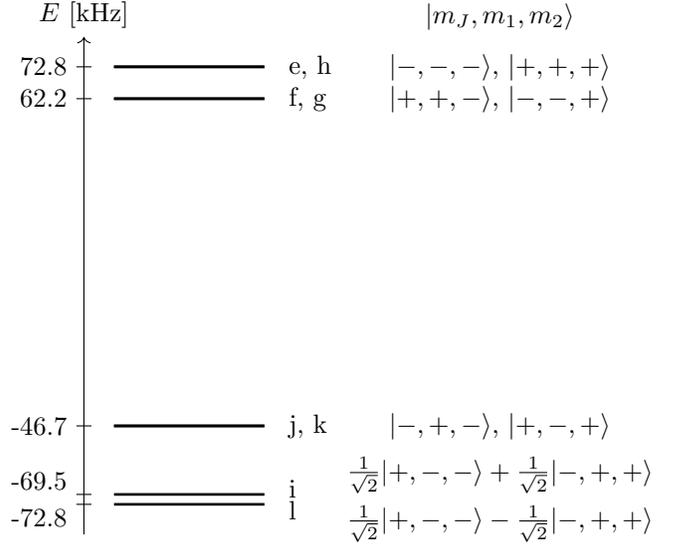
\begin{figure}
	\centering
	\begin{tikzpicture}[scale=4]
  \def\scaling{0.01}
  \def\eE{72.8}
  \def\hE{72.8}
  \def\fE{62.2}
  \def\gE{62.2}
  \def\jE{-46.7}
  \def\kE{-46.7}
  \def\iE{-69.5}
  \def\lE{-72.8}

  \def\xs{0.1}
  \def\linelen{0.5}
  \def\lwidth{1}
  \draw [line width = \lwidth] (\xs,\eE*\scaling) -- (\xs+\linelen,\eE*\scaling);
  \draw [line width = \lwidth] (\xs,\hE*\scaling) -- (\xs+\linelen,\hE*\scaling);

  \draw [line width = \lwidth] (\xs,\fE*\scaling) -- (\xs+\linelen,\fE*\scaling);
  \draw [line width = \lwidth] (\xs,\gE*\scaling) -- (\xs+\linelen,\gE*\scaling);

  \draw [line width = \lwidth] (\xs,\jE*\scaling) -- (\xs+\linelen,\jE*\scaling);
  \draw [line width = \lwidth] (\xs,\kE*\scaling) -- (\xs+\linelen,\kE*\scaling);

  \draw [line width = \lwidth] (\xs,\iE*\scaling) -- (\xs+\linelen,\iE*\scaling);
  \draw [line width = \lwidth] (\xs,\lE*\scaling) -- (\xs+\linelen,\lE*\scaling);

  \def\roffset{0.05}
  \def\yoffset{0.02}
  \node[align=left, anchor = west] (eh) at (\xs+\linelen+\roffset,\eE*\scaling) {e, h};
  \node[align=left, anchor = west] (fg) at (\xs+\linelen+\roffset,\fE*\scaling) {f, g};
  \node[align=left, anchor = west] (jk) at (\xs+\linelen+\roffset,\jE*\scaling) {j, k};
  \node[align=left, anchor = west] (i) at (\xs+\linelen+\roffset,\iE*\scaling+\yoffset) {i};
  \node[align=left, anchor = west] (l) at (\xs+\linelen+\roffset,\lE*\scaling-\yoffset) {l};

  \def\yoffset{0.1}
  \def\len{0.05}
  \draw[->] (0, \lE*\scaling-\yoffset) -- (0, \eE*\scaling+\yoffset) node[above] {$E$ [kHz]};

  \draw (0-\len/2, \eE*\scaling) node[anchor=east,left] {\eE}
     -- (\len/2,\eE*\scaling);
   \draw (0-\len/2, \fE*\scaling) node[anchor=east,left] {\fE}
      -- (\len/2,\fE*\scaling);
  \draw (0-\len/2, \jE*\scaling) node[anchor=east,left] {\jE}
     -- (\len/2,\jE*\scaling);
  \draw (0-\len/2, \iE*\scaling) node[anchor=south east,left,yshift = 5] {\iE}
    -- (\len/2,\iE*\scaling);
  \draw (0-\len/2, \lE*\scaling) node[anchor=north east,left,yshift = -5] {\lE}
    -- (\len/2,\lE*\scaling);

  \def\xoffset{1}
  \def\yshift{5}
  \node[right=\xoffset of eh, left, anchor=west, yshift = 19] (ketlabel) {$\ket{m_J,m_1,m_2}$};
  \node at (eh -| ketlabel) {$\ket{-,-,-}$, $\ket{+,+,+}$};
  \node at (fg -| ketlabel) {$\ket{+,+,-}$, $\ket{-,-,+}$};
  \node at (jk -| ketlabel) {$\ket{-,+,-}$, $\ket{+,-,+}$};

  \node[yshift = \yshift] at (i -| ketlabel) {$\frac{1}{\sqrt{2}}\ket{+,-,-}+\frac{1}{\sqrt{2}}\ket{-,+,+}$};
  \node[yshift = -\yshift] at (l -| ketlabel) {$\frac{1}{\sqrt{2}}\ket{+,-,-}-\frac{1}{\sqrt{2}}\ket{-,+,+}$};

\end{tikzpicture}
	\caption{Hyperfine level structure of TlF $\ket{J=1,\ m_J=\pm1}$ states in $\mathcal{E}=30\kVcm$ and $\Bsca = 0$, as will be present in the Main Interaction Region.  Lettered naming labels correspond to those in \cite{wilkening1984search}. For brevity, only the sign of the quantum numbers is shown; the full values are $m_J =\pm 1$, $m_1=\pm \sfrac{1}{2}$, $m_2=\pm\sfrac{1}{2}$, where $m_1 (m_2)$ is the Tl (F) nuclear spin. The zero of energy is arbitrary.}
	\label{fig:levels_interaction_region}
\end{figure}
In order to measure the Schiff moment, a large, uniform external electric field ($\Esca_\mathrm{MI}=30\kVcm$, along $\hat{z}$) polarizes the molecules. In this large field, energies of the $\widetilde{J}=1,m_J=\pm 1$ manifold of hyperfine states are close, relative to their splittings to all other $\widetilde{J},m_J$ states. The states within the $\widetilde{J}=1,m_J=\pm 1$ subspace are then well-described \cite{cho1991search} by the effective Hamiltonian\footnotemark
\begin{equation}
    \begin{split}
        \mathcal{H}_\textrm{eff} =& -\mu_J J_z \Bsca_z + (-\mu_1 \vec{I}_1 -\mu_2 \vec{I}_2)\cdot \Bvec  \\
        &- \mu_1 I_{1z} J_z \Bsca^{\rm int}_1 - \mu_2 I_{2z} J_z \Bsca^{\rm int}_2 + C_{\rm s} I_{1z} I_{2z} \\
        &+ C_{\rm t} (J_+^2I_{-,1}I_{-,2} + h.c.) + W_S\,S \frac{\vec{I}_1}{I_1}\cdot\hat{\vec{n}}.
    \end{split}
\end{equation}
Here,  $\Bsca^{\rm int}$ denotes an effective intra-molecular magnetic field along $\langle \vec{J} \rangle$ that arises from the spin-rotation terms in Eq. \ref{eq:hyperfine_hamiltonian}, while $C_{\rm s}$ and $C_{\rm t}$ are effective scalar and tensor spin-spin interactions, respectively, that arise from the spin-spin terms in Eq. \ref{eq:hyperfine_hamiltonian}.  The subscripts $z,+,-$ on operators refer to the usual angular momentum projection, raising, and lowering operators, respectively.
\footnotetext{The effective Hamiltonian formulation (see, e.g., Ref.~\cite{brown2003rotational}) is useful to describe states in a near-degenerate subspace that couple only weakly to all other states in a larger Hilbert space. We split the Hamiltonian of Eq.~\ref{eq:hyperfine_hamiltonian} into a large 0$^{\rm th}$-order term, $\mathcal{H}^{(0)} = \mathcal{H}_{\rm rot} + \mathcal{H}_{\rm S}$, and a small 1$^{\rm st}$ order term, $\mathcal{H}^{(1)} = \mathcal{H}_{\rm Z}+\mathcal{H}_{\rm sr} + \mathcal{H}_{\rm ss}$. We then compute eigenstates of $\mathcal{H}^{(0)}$, operating on the full Hilbert space of all spin-rotation levels, and use the set of all $\widetilde{J}=1,m_J=\pm1$ states as (degenerate) basis states for the subspace.
To lowest order, the effective Hamiltonian $\mathcal{H}_\textrm{eff}$ acting on this subspace consists only of the terms in $\mathcal{H}^{(1)}$ that couple states within the subspace (including diagonal terms). That is, $\mathcal{H}_\textrm{eff}$ is derived by first discarding all operator terms in the full Hamiltonian of Eq. \ref{eq:hyperfine_hamiltonian} that change $m_J$ by $\pm 1$ unit, and hence only connect states in the subspace to distant outside states, then computing matrix elements of the remaining operators between the basis states. The resulting coefficients $C_{\rm s}$ and $C_{\rm t}$ are linear combinations of $c_3$ and $c_4$, but, because $J$ is not a good quantum number, simple analytic expressions cannot be given for their weights.}

The degree of electric polarization, $\mathcal{P}$, is given by
\begin{equation}
    \mathcal{P} = \langle \hat{\vec{n}}\cdot \hat{\mathcal{E}}\rangle = \langle \cos\theta \rangle,
\end{equation}
where $\theta$ is the angle between $\hat{n}$ and 
$\Evec_{\rm MI}$. For a state in the $J=1,\ m_J=\pm 1$ manifold at $\Esca_\mathrm{MI} = 30 \kVcm$, $\mathcal{P} = 0.547$.  
To determine the $^{205}$Tl NSM, we measure how the energy splitting between two states with the same $m_J$ values but opposite Tl spin projections ($m_1 = \pm 1/2$) changes, when $\Evec_{\rm MI}$ is reversed.
As discussed in Sec.~\ref{sec:introduction}, when $S$ is non-zero this splitting will shift by $\pm 2\Delta_{\rm CPV} = \pm 2W_S\,S\,\mathcal{P}\,\textrm{sgn}(\Esca_{\rm MI})$ (Fig.~\ref{fig:edm_shift}). 
The states in the $J=1, m_J=\pm 1$ manifold in the electric field of magnitude $\Esca_{\rm MI}$ are shown in Fig.~\ref{fig:levels_interaction_region}; the pairs that flip only the Tl spin $\vec{I}_1$ are j/e and k/h, both with a separation of $119.517\,$kHz due mainly to the effective internal magnetic field and the scalar spin-spin interaction.  Hence, we seek to measure the splittings between these levels, and how they change when $\Evec$ or other experimental parameters are reversed.

The energy splitting is measured with a classic SOF technique for nuclear magnetic resonance \cite{PhysRev.78.695,ramsey1951phase}, albeit with the external $\Bsca$-field set to zero. 
The RF drive frequency is set to resonance with the j/e or k/h transition, i.e., $\hbar\omega_\mathrm{RF} \approx \mu_\mathrm{Tl}\mathcal{B}_\mathrm{int}- C_s/2$.
The first RF pulse in the SOF sequence creates a superposition of the Tl spin-up and spin-down states; e.g. starting in state e, the $\pi/2$ pulse creates a superposition between states e and j. Then, during a period $T$ of free precession, the accumulated phase between the up (e or h) and down (j or k) states is
\begin{equation}
    \begin{split}
        \phi \approx & \left( -\mu_\textrm{Tl}\mathcal{B}_\textrm{int}\mathrm{sgn}(m_J) + C_s m_{I_2} \right. \\ & \left. + 2 W_S\,S\,\mathcal{P}\mathrm{sgn}(\Esca_{\rm MI}) \right)T/\hbar.
    \end{split}
\end{equation}
The second RF pulse maps $\phi$ onto the relative population in the spin-up and spin-down states.
The probability of a transition from spin-up to spin-down is \cite{ramsey1951phase}
\begin{equation}
    P_{\uparrow \rightarrow \downarrow} = \sin^2 \frac{1}{2} \Omega_{\rm RF} \tau \cos^2 \frac{1}{2}\left(\phi_{\rm CPV} + \phi_\mathrm{SOF}\right),
\end{equation}
where $\phi_\mathrm{SOF}$ is the phase offset between the first and second RF pulses, $\Omega_{\rm RF}$ is the Rabi frequency of the RF magnetic field, $\tau$ is the time spent in the perturbing RF field, and
\begin{equation}
    \phi_{\rm CPV} = 2W_S\,S\,\mathcal{P}\,\mathrm{sgn}(\Esca_{\rm MI})\,T/\hbar = 2\,\Delta_{\rm CPV}\,T/\hbar.
\end{equation}
The aim is to determine $\phi_{\rm CPV}$; from its measured value and the known value of $T$, $\Delta_{\rm CPV}$ can be found.
The phase difference $\phi_\mathrm{SOF}$ will be set to have magnitude $\pi/2$ and to alternate in sign: $\phi_\mathrm{SOF} = \pm\sfrac{\pi}{2}$. This yields maximal sensitivity to the small energy shift $\Delta_{\rm CPV}$.

In a zero magnetic field environment with $\Evec_\mathrm{MI}$ present, $\phi_{\rm CPV}$ only accumulates due to the $T$-violating frequency shift. In practice, magnetic fields cannot be fully eliminated from the interaction region, and will generate additional frequency shifts. To minimize the contribution of magnetic fields, we will construct a magnetic shield consisting of several concentric cylinders. Currently, a 4-cylinder shield, with 12 layers of Metglas high-permeability material \cite{jiles2015magnetism} on each cylinder, is planned. These will be augmented with shim coils inside and outside the shields for additional magnetic field control. We aim to achieve sub-$10\,\mu$G residual fields (see Sec.~\ref{sec:frequency_shifts}).

The externally applied electric field $\Evec_\mathrm{MI}$ will be generated with two quartz electrodes of $3\,$m length, separated by $2\cm$, with a Rogowski profile \cite{rogowski1923profile, highvoltage2000} to prevent formation of large edge fields. The electrodes will be coated with a conductive water-based colloidal graphite coating. The choice of coated quartz electrodes was made to minimize electrical conductance that leads to magnetic Johnson noise \cite{PhysRevA.60.1717, doi:10.1063/1.2737357, doi:10.1063/1.2885711}. To further minimize Johnson noise, the vacuum chamber surrounding the MI region will be constructed from a quartz tube ($3.5\,$m long, $26\cm$ O.D., $2\cm$ wall thickness). To prevent charge buildup, the inner surface of the tube will be grounded with either a thin conducting sheet or a conductive coating. 

Coils to produce the SOF NMR fields will be placed outside the vacuum chamber, placed symmetrically about the center of the electrodes and separated by distance $L_{\rm SOF} \approx 2.5$ m.  Several additional coils will be mounted to allow application of small DC $\Bvec$-fields and gradients in various directions; these will be used mostly for diagnosing and correcting systematic errors.  

\subsection{State Preparation Region C}
\label{sec:state_preparation_region_c}
After the main interaction region, molecules populate two states, $\ket{J=1,m_J,m_{I_1}=+1/2,m_{I_2}}$ and $\ket{J=1,m_J,m_{I_1}=-1/2,m_{I_2}}$, which cannot be distinguished optically. To enable optical detection, population from one or both of these states will be transferred to different rotational states, which are spectrally resolvable by a probing laser.  This will allow optical detection of each original spin-state population (Sec.~\ref{sec:detection_region}). An optimized scheme for this state transfer mechanism is currently being investigated, but will likely involve a microwave $\pi$-pulse, mirroring state preparation region B.

\subsection{Fluorescence Detection Region}
\label{sec:detection_region}
Detecting the populations in the two rotational states resulting from state transfer region C, which carry the information about the accumulated phase $\phi$, will be achieved with optical cycling to maximize the number of emitted photons from each molecule. This cycling, which has been demonstrated experimentally in TlF \cite{privateHunter2019}, will allow for near unit-efficiency detection of each molecule. 

The rotational sublevels of the TlF ground state $X^1\Sigma^+$ are far enough apart to require two detection lasers. Rapid switching between the lasers will allow for quasi\hyp{}simultaneous readout of both the spin-up and spin-down populations in a single molecular-beam pulse, minimizing the effect of molecule number fluctuations within and between pulses \cite{kirilov2013shotnoise}. The switching, to be accomplished with acousto-optic modulators, will allow enough dead time between switches for the excited state to decay, but also will be rapid enough such that each molecule sees both laser frequencies multiple times while traveling through the optical interaction region. A similar scheme is implemented by the ACME experiment \cite{andreev_improved_2018}. 

The resulting fluorescence will be collected by a combination of high numerical aperture lenses and mirrors to cover a total solid angle of $\approx 0.3\! \times\! 4\pi$\,sr.  With PMT quantum efficiency of $\approx 25\%$, each emitted photon will then be detected with $\approx 7.5\% $ efficiency.  Hence, scattering $\gtrsim 30$ photons per molecule will be sufficient that each molecule is detected with $\gtrsim 90\%$ probability. Based on known branching ratios for decay out of each cycling transition in TlF \cite{norrgard2017hyperfine}, this should be feasible.

The fluorescence signals $\mathrm{S}_\uparrow$ and $\mathrm{S}_\downarrow$, corresponding to populations in the Tl spin-up and spin-down states after the SOF sequence, are then used to compute the asymmetry $\mathcal{A}$, defined as: 
\begin{equation}
    \mathcal{A} \equiv \frac{\mathrm{S}_\uparrow - \mathrm{S}_\downarrow}{\mathrm{S}_\uparrow + \mathrm{S}_\downarrow}.
\end{equation}
With the SOF drive frequency on resonance,  
\begin{equation}
    \mathcal{A} \approx 1 - 2\sin^2 \Omega_\text{RF}\tau\cos^2 \frac{1}{2}\left(\phi_{\rm CPV} + \phi_\mathrm{SOF}\right).
\end{equation}
For $\phi_\mathrm{SOF} = \pm \sfrac{\pi}{2}$ and $\Omega_\text{RF}\tau=\pi/2$, this simplifies to $\mathcal{A} \approx \pm \sin\phi_{\rm CPV} \approx \pm \phi_{\rm CPV}$.

\subsection{Laser and Microwave Systems}

To accomplish the Schiff moment measurement in TlF, three UV lasers at $272\nm$ are required:  one for rotational cooling, and two for the quasi-simultaneous state readout.

Three IR seed lasers are frequency doubled twice to reach UV.  The IR seeds are tunable fiber lasers, providing $\sim10\mW$ per laser at $1087\nm$.  All of the seed beams are amplified with Yb fiber amplifiers, then doubled into green ($544\nm$). Two of the systems accomplish this by coupling their amplified output ($1.4\,$W) to a home-built resonant cavity containing a PPKTP crystal, delivering $\sim 500\mW$ of green light in free space.  A third system uses a high-power amplifier and a single-pass doubler to produce $1.4\,$W of green light from a single-mode fiber. For all systems, the green light is coupled into a commercial resonant cavity containing a BBO crystal. This frequency-doubles the green light to produce $\lesssim 80\mW$ of $272\nm$ single frequency, tunable light from each system.

These UV lasers are locked to a tunable offset from resonance, using a frequency transfer scheme employing scanning confocal cavities and a single stable reference laser~\cite{lindsay1991scanningcavity,jaffe1993scanningcavity,zhao1998scanningcavity,rossi2002scanningcavity}.  For the reference laser, we tightly lock a tunable external-cavity diode laser to a $D_2$ transition in atomic Cs, using modulation transfer spectroscopy (MTS) \cite{zi2017laser}, to achieve absolute frequency stability of better than 100 kHz.

Several microwave-frequency $\Evec$-fields are required to couple the rotational ground states in the state preparation regions. To control the spatial distribution of these microwave fields, we use microwave quasi-optical spot-focusing horns that create nominally Gaussian, traveling-wave beams, with their waists centered on the molecular beam.  The microwave beams enter and exit the vacuum chamber through windows large enough to ensure negligible clipping of the intensity profile. These windows have a thickness $\lambda/2$ that, much like an anti-reflection coating for optical windows, causes destructive interference between reflections off the front and back surfaces of each window. Furthermore, each beam, on exiting the chamber, is received by a horn identical to the transmitting horn, at the same distance from the waist. Hence, the microwave beam is nominally matched in spatial mode to the receiving horn, which is terminated in 50 Ohms to absorb the incident power. These measures minimize the reflected intensity, to prevent unwanted standing wave components. Each horn is fed through an orthomode transducer; switches after the microwave generators and amplifiers can direct the full power to either input port, and hence deliver either allowed linear polarization to the molecules.

\subsection{State Evolution Between Regions}

The different functional regions of \CENTREX\ require $\Evec$- and $\Bvec$-fields of widely varying magnitude and orientation. Hence, in the spaces between the functional regions, spatially-varying fields will be present. These manifest as time-varying fields in the molecules' rest frame, resulting in unwanted transfer of molecular population from the desired state to undesired states. This loss reduces statistical sensitivity and can lead to systematic errors. For example, if molecular population is lost non-uniformly over the cross section of the molecular beam, an inhomogeneous distribution of molecules will result. When combined with spatial field gradients within the Main Interaction region, this has been observed to cause systematic errors in related experiments \cite{andreev_improved_2018}. Understanding how the relevant quantum states evolve when molecules travel between functional regions is therefore important both in terms of optimizing the statistics and avoiding systematic errors in \CENTREX.

We have performed extensive numerical simulations to identify optimized schemes for transfer between regions in \CENTREX, and to understand how undesired states can be populated here.  We find it should be possible to achieve near 100\% efficiency in all cases.  Because the between-region state evolution in \CENTREX\ is non-trivial to understand, yet appears to be sufficiently under control, we do not discuss it in detail here.  Interested readers can find a thorough treatment in Appendix A.

\section{Sensitivity and Systematics}
\label{sec:sensitivity_and_systematics}

\subsection{Anticipated Sensitivity}

The molecule-shot-noise limited (SNL) sensitivity for a SOF frequency measurement in a beam is given by
\begin{equation}
    \delta \nu_\mathrm{SNL} = \frac{1}{2\pi T}\frac{1}{C_\mathrm{SOF}}\frac{1}{\sqrt{N_d N_p}} Z.
\end{equation}
Here $T$ is the total interaction time in the MI region, $L_{\rm SOF}/\langle v_z\rangle$,
$C_\mathrm{SOF}$ is the SOF fringe constrast, $N_d$ is the number of molecules detected per beam pulse, and $N_p$ is the number of pulses used in the measurement. \CENTREX\ is expected to achieve $C_{\rm SOF} \approx 1$, as in the ACME electron EDM measurement that uses a similar detection scheme \cite{andreev_improved_2018}.  The factor $Z$, which takes values $1\! <\! Z\! < \!\sqrt{2}$, accounts for excess noise that can arise when detecting fluorescence from a partially-closed cycling transition \cite{PhysRevA.98.053823}.  To be conservative, we take $Z=\sqrt{2}$.

We estimate $N_d$ as follows.  The measured time\hyph averaged beam intensity is $5\times 10^{12}\,$molecules/state/sr/s (Sec.\ \ref{sec:beamsource}), corresponding to $1\times 10^{11}\,$molecules/state/sr/pulse. Rotational cooling results in a simulated 24-fold increase in the number of molecules in the desired $F=0,m_F=0$ state (Sec.\ \ref{sec:rotational_cooling}).
Combining the simulated state transfer efficiencies of SPA (99\%), SPB (96\%), and SPC (96\%) (Secs.\ \ref{sec:state_preparation_region_a}, \ref{sec:state_preparation_B} and \ref{sec:state_preparation_region_c}), giving a cumulative transfer efficiency of $\approx 91\%$.  Given the distance of the FD region from the source and its transverse area 18 mm $\times$ 30 mm, the solid angle subtended by the FD region is $1.3\times 10^{-5}\,$sr. From simulations of the EQL (Sec.\ \ref{sec:electrostatic_lens}), the gain in signal from energizing the lens should be 24. Combining these with the anticipated detection efficiency of 90\%, we expect $N_d \approx 6.1\ee{8}\,$molecules/pulse to be detected in the FD region.

Combined with $\langle v_z \rangle = 184\,$m/s and $L_{\rm SOF} \approx 2.5\,$ m, we estimate a shot noise-limited frequency shift sensitivity of 
\begin{equation}
    \label{eq:shot_noise}
	\delta \nu_\mathrm{SNL} \approx \frac{0.7}{\sqrt{N_p}}\mHz.
\end{equation}
With a total measurement time of 300 hours and a $50\,$Hz repetition rate, corresponding to $N_p \approx 5.4\ee{7}$, the final sensitivity is projected to be $\delta \nu_\mathrm{SNL} \approx 90\,$nHz. Recalling that the $CPV$ energy shift is $2\Delta_{\rm CPV}$, this is equivalent to $\delta \Delta_\mathrm{CPV} \approx 45$ nHz.
For comparison, the previous best limit achieved $\delta\Delta_\mathrm{CPV} \approx 120~\mu$Hz \cite{cho1991search}. Hence we anticipate that \CENTREX\ can achieve a 2500-fold statistical improvement over the CPV limits given in Eq.~\ref{eq:prev_best_lims}.

\subsection{Extracting \texorpdfstring{$\Delta_{\rm CPV}$}{dCVP}}

We will extract the CPV energy shift $\Delta_{\rm CPV}$ from our data using schemes similar to those used in prior experiments \cite{wilkening1984search,cho1991tight,regan2002new,andreev_improved_2018} and described briefly here.

Recall that under ideal conditions, the signal asymmetry is given by 
$\mathcal{A}= \sgn\left(\phi_\mathrm{SOF}\right) \sin \phi_{\rm CPV}$.
In practice, various experimental imperfections (e.g. deviations from exact RF phase and/or resonance frequency) generate an additional accumulated phase $\phi'$ during free precession in the interaction region. 
This modifies the asymmetry, such that $\mathcal{A}= \sgn\left(\phi_\mathrm{SOF}\right) \sin (\phi_{\rm CPV} + \phi')$.
To isolate the CPV phase term, we measure $\mathcal{A}$ under two different conditions where the sign of $\Delta_\mathrm{CPV}$ reverses. This is the case, for example, when the direction of $\Evec_\mathrm{MI}$
is reversed.
Then we calculate
\begin{equation}
    \mathcal{A}_{+\Esca_\mathrm{MI}} - \mathcal{A}_{-\Esca_\mathrm{MI}} \approx \pm 2\phi_{\rm CPV},
\end{equation}   
independent of $\phi'$ so long as $\phi' \ll 1$.  We refer to the reversal of $\Evec_\mathrm{MI}$ as $E$-modulation, and assign the parameter $E = \pm 1 = \mathrm{sgn}(\Evec_\mathrm{MI} \cdot \hat{z})$.

It is possible to reverse the sign $\Delta_\mathrm{CPV}$ in several other ways as well. For example, simultaneously reversing the magnetic fields in state preparation regions B and C, $\Bvec_\mathrm{SPB}$ and $\Bvec_\mathrm{SPC}$, reverses the signs of all the angular momenta relative to the fixed laboratory $z$-axis. This corresponds to changing the signs of $m_{I1}, m_{I2}$, and $m_J$, and hence also the sign of $\Delta_{\rm CPV}$. We refer to this reversal as $B$ modulation and define $B = \mathrm{sgn}(\Bvec_\mathrm{SPB} \cdot \hat{z})$. Finally, changing the frequency of the microwave fields in SPB and SPC makes it possible to select which states to initially populate for use in the MI region. For example, the transitions e$\leftrightarrow$j and h$\leftrightarrow$k are time-reversed versions of each other, meaning the effective internal magnetic field has opposite sign between them. This also changes the sign of $\Delta_{\rm CPV}$.
We refer to this reversal as $M$ modulation, and define $M = \pm 1$ corresponding to the h$\leftrightarrow$k and e$\leftrightarrow$j transitions, respectively.

While any of these modulations will, in principle, serve to isolate the contribution of $\phi_\mathrm{CPV}$, in practice we will employ all of them to provide various in-situ diagnostics. Our parameter naming convention here follows that of Ref.~\cite{cho1991search}. 

It proves useful to also employ a few more modulations. In particular, modulating the sign of the phase offset $\phi_{\rm SOF}$ between the RF coils, as discussed in Sec.~\ref{sec:interaction_region}, changes the sign of the asymmetry $\mathcal{A}$. This $P$ modulation has no effect on $\Delta_{\rm CPV}$.

The fringe contrast $C_\mathrm{SOF}$ can be measured by alternatingly offsetting the SOF drive frequency from its resonance value ($f_0 = 119.516\kHz$) by $\pm f_F$, where $f_F$ is small compared to the SOF NMR linewidth. The quantity $\mathcal{A}_{+f_F}-\mathcal{A}_{-f_F}$ determines the slope of the resonance and hence $C_\mathrm{SOF}$. This $F$ modulation has no effect on $\Delta_{\rm CPV}$.

During the NSM measurement, all these modulation parameters will be frequently switched to determine $\Delta_\mathrm{CPV}$ and to diagnose various possible contributions to $\phi'$. 

We denote the various combinations of asymmetries (i.e., phases) that can be constructed from these modulations with the notation $\mathcal{S}_{p1,p2,...}$. Here, the subscripts denote a linear combination of phases odd under the listed modulation parameters $p1,p2,...$ and even under all other modulations. For example, the total phase shift $\phi_{\rm CPV}$ is determined via
\begin{equation}
    \phi_\mathrm{CPV} \propto \mathcal{S}_{PEBM} = \sum_i\left(P_i\,E_i\,B_i\,M_i\right)\mathcal{A}_i,
\end{equation}
where $P_i,\,E_i,\,B_i,\,$ and $M_i$ are the signs of the modulation parameters during the $i^\mathrm{th}$ dataset, and $\mathcal{A}_i$ is the measured asymmetry for that dataset.  
The quantity
\begin{equation}
    \mathcal{S}_\mathrm{PF} = \sum_i \left(P_i\,F_i\right)\mathcal{A}_i,
\end{equation}
which determines the slope of the frequency vs.~phase curve (and hence also the fringe contrast $C_\mathrm{SOF}$), is used to convert $\phi_\mathrm{CPV}$ to frequency units.

\subsection{Known Systematic Errors}
\label{sec:frequency_shifts}

\begin{table*}
	\normalsize
	\makebox[\textwidth][c]{%
	    \resizebox{\textwidth}{!}{
		\begin{tabu} to \textwidth {@{}r @{\hspace{3pt}} ccc cc ccc cc ccc cc@{}}
			\toprule
			
			& \multicolumn{4}{c}{State 1} & \multicolumn{4}{c}{State 2} & $f_0$ & $df_0/d\mathcal{B}_z$ & $df_0/d\mathcal{E}_z$ & shift $\mathcal{B}_\mathrm{mot}$ & $S$ & \multicolumn{2}{c}{$|\bra{\cdot}\mathcal{H}_\text{Z}\ket{\cdot}|$ [kHz]\hspace{1.5ex}\mbox{}}\\
			\cmidrule(r{2.0ex}){2-5}\cmidrule(r{2.0ex}){6-9}\cmidrule(r{2.0ex}){15-16}
			What flips?\hspace{1ex} & $l$ & $m_J$ & $m_{I_1}$ & $m_{I_2}$ & $l$ & $m_J$ & $m_{I_1}$ & $m_{I_2}$ & kHz & [mHz/$\mu$G] & [mHz/(V/cm)] & [mHz/$\mu$G] & - & x,y & z\\
			\midrule
			
			\multirow{2}{*}{$m_{I_1}\;\Big\{$}             & e & $-$ & $-$ & $-$ & j & $-$ & $+$ & $-$ & \multirow{2}{*}{119.52} & $+2.49$ &  \multirow{2}{*}{$- 31.50$} & $+4.66\ee{-5}$ & \multirow{2}{*}{0.95} & \multirow{2}{*}{1.33} & \multirow{2}{*}{0.00} \\
			& h & $+$ & $+$ & $+$ & k & $+$ & $-$ & $+$ &  & $-2.49$                &                       &  $+5.22\ee{-5}$ & & & \\

			\dashedrule
			
			\multirow{2}{*}{$m_{I_1},m_{I_2}\;\Big\{$}             & f & $+$ & $+$ & $-$ & k & $+$ & $-$ & $+$ & \multirow{2}{*}{108.92} &  $+1.52$ &  \multirow{2}{*}{$- 3.57$} & $-1.17\ee{-4}$ & \multirow{2}{*}{0.99} & \multirow{2}{*}{0.00} & \multirow{2}{*}{0.09}\\
			& g & $-$ & $-$ & $+$ & j & $-$ & $+$ & $-$  &               & $-1.52$                &                       &  $-1.23\ee{-4}$ & & &\\
	
			\dashedrule

			\multirow{2}{*}{$m_{I_2}\;\Big\{$}             &e & $-$ & $-$ & $-$ & g & $-$ & $-$ & $+$ & \multirow{2}{*}{10.59} & $+4.00$ &  \multirow{2}{*}{$- 27.93$} & \multirow{2}{*}{$+1.69\ee{-4}$} & \multirow{2}{*}{0.04} & \multirow{2}{*}{1.88} & \multirow{2}{*}{0.00} \\
			& h & $+$ & $+$ & $+$ & f & $+$ & $+$ & $-$ & &              $-4.00$  &               &  & & &\\

			\bottomrule
	\end{tabu}}}
	\normalsize
	\caption{All non-degenerate pairs of $|m_J|=1$ states in the $\widetilde{J}=1$ manifold at $\Esca_\mathrm{MI} = 30 \kVcm$ that do not involve the states $\ket{\text{i}}$ and $\ket{\text{l}}$ or a flip of $m_J$. The quantum numbers given are that of the largest decoupled-basis component. State labels $l$ are as in \cite{wilkening1984search} and Fig.~\ref{fig:levels_interaction_region}. The quantities $df_0/d\Bsca_z$ and $df_0/d\Esca_z$ give the slope of the resonance frequency with respect to the external magnetic field and electric field, respectively. Shift $\mathcal{B}_\mathrm{mot}$ indicates the resonance frequency shift, due to the motional field that accompanies $\mathcal{E}$ reversal, with respect to a stray field component $\Bsca_y$. $S$ denotes the sensitivity to the NSM relative to the maximum possible value; it is given by $\left|\langle I_{1,z}\rangle_1-\langle I_{1,z}\rangle_2\right|$ for the transition between states 1 and 2. $f_0$ indicates the transition frequency between the two states. $|\bra{\cdot}\mathcal{H}_\text{Z}\ket{\cdot}|$ indicates the magnitude of the transition dipole moment between states 1 and 2. All shifts are calculated from diagonalization of the ground-state Hamiltonian (Eq.~\ref{eq:hyperfine_hamiltonian}).}
	\label{tab:freq_shifts}
\end{table*}

Here we discuss the anticipated magnitude of some known systematic errors in \CENTREX. Our discussion closely follows the notation and analysis of Ref.~\cite{cho1991search}.

\subsubsection{Imperfect \texorpdfstring{$\mathcal{E}$}{E}-field Reversal}
The separation between the Tl spin up/down states in the $J=1,\,m_J=\pm1$ manifold changes slightly when the externally applied $\Esca$-field changes in magnitude.\footnote{This is due to 2nd-order spin-spin and spin-rotation couplings to distant $\ket{\widetilde{J},m_J}$ states.} Any non-reversing contribution to $\Evec_\mathrm{MI}$, e.g., from a stray DC field, thus leads to a  frequency shift in the NMR transition that changes with the orientation of $\Evec_\mathrm{MI}$. By brute-force diagonalization of the ground-state Hamiltonian of Eq.~\ref{eq:hyperfine_hamiltonian} with $\Esca = 30\kVcm$, the frequency shift can be calculated: see Tab.~\ref{tab:freq_shifts}. The pairs of states ej and hk that are used for the measurement both have an identical shift of $-31.5\,$mHz/(V/cm).  Assuming the non-reversing $\Evec$-field component does not change significantly between subsequent $M$ and $B$ reversals, this effect will be suppressed in the quantity $\mathcal{S}_{EBMP} \propto \phi_\mathrm{CPV}$ that is odd under both $M$ and $B$. For the residual shift to be below our anticipated sensitivity, we will require a small non-reversing $\Evec$-field as well as accurate changes of both $\Bsca$ and the initial state of the NMR transition. The former quantity can be determined from the signal combination $\mathcal{S}_{EP}$, and then nulled by applying an appropriate offset voltage; the inaccuracy in latter two can be determined from $\mathcal{S}_{EBMP}$ when a deliberately large non-reversing $\Evec$-field is applied, then nulled if necessary. 

\subsubsection{Stray \texorpdfstring{$\mathcal{B}$}{B}-Fields}

The \CENTREX\ measurement will be performed with a nominally zero $\Bvec$-field in the Main Interaction region. Significant effort will be made to minimize any residual stray fields, but nevertheless some will persist. These can arise, e.g., from leakage through, or residual magnetization of, the magnetic shielding. These stray $\Bsca$-fields can lead to systematic errors via two mechanisms: direct shifts, and in combination with motional-field effects.

For the pair of states ej and hk, a $\Bvec$-field along $\Evec_\mathrm{MI}$ (i.e., $\Bsca_z$) generates a direct frequency shift of $\pm2.5~\mHz$/$\mu$G, where the sign applies for the ej and hk transition, respectively. \CENTREX\ aims for sub--$10\,\mu$G residual $\Bsca$-fields, which will alone shift the transition frequency by $\mathcal{S}_\mathrm{B} \approx \pm 25\mHz$.  

Consider the total effective magnetic field $\mathcal{B}_\mathrm{MI}$ in the MI region. This field has contributions from several physical mechanisms; we write
\begin{equation}
    \mathcal{B}_\mathrm{MI} = \mathcal{B}_\mathrm{int} + \mathcal{B}_\mathrm{st} + \mathcal{B}_\mathrm{SP} +
    \mathcal{B}_\mathrm{LC},
\end{equation}
where $\mathcal{B}_\mathrm{int}$ is the intra-molecular magnetic field, $\mathcal{B}_\mathrm{st}$ is a static stray field, $\mathcal{B}_\mathrm{SP}$ is from the magnetic fields in SPB and SPC penetrating into the MI region, and $\mathcal{B}_\mathrm{LC}$ is from leakage currents in the electrode structure.
Both $\mathcal{B}_\mathrm{int}$ and $\mathcal{B}_\mathrm{SP}$ change sign under $B$ modulation. Under $M$ modulation, only $\mathcal{B}_\mathrm{int}$ changes sign. We do not expect $\mathcal{B}_\mathrm{st}$ to change significantly under any of the modulations.
So, in order to fully suppress the direct shifts due to stray magnetic fields, all three modulations $E$, $B$ and $M$ are required.

However, none of these modulations help to distinguish Zeeman shifts due to $\mathcal{B}_\mathrm{LC}$ from a true NSM signal, since both reverse under $E$ modulation.  Hence, as usual for EDM experiments, it will be very important to minimize the leakage current $I_\mathrm{LC}$.  Using the standard crude approximation for a worst-case scenario of $\Bsca_\mathrm{LC}$ (where all leakage current flows around a helical path between electrodes), we find that $I_\mathrm{LC}$ could conceivably need to be as low as $\sim\!1$~nA to absolutely ensure that this systematic error is less than our anticipated statistical sensitivity. Because this may prove challenging, we discuss possible methods to reduce our sensitivity to leakage currents in Sec. \ref{Sec:InternalComagnetometry}.

The magnitude of all other contributions to $\mathcal{B}_\mathrm{MI}$ can be determined from appropriate signal combinations. For example, $\mathcal{S}_\mathrm{BMP} \approx 5\mathcal{B}_\mathrm{st}\mHz/\mu$G determines $\mathcal{B}_\mathrm{st}$, since $B$ and $M$ work together to reverse $\mathcal{B}_\mathrm{int}$ but keep the orientation of the spins. Similarly, $\mathcal{S}_\mathrm{MP} \approx 5\mathcal{B}_\mathrm{SP}\mHz/\mu$G determines $\mathcal{B}_\mathrm{SP}$,  since $M$ flips the direction of the spins relative to the fields in SPB and SPC regions. Once measured, these fields can be nulled; then, by deliberately exaggerating each component separately, their residual effects on $\Delta_\mathrm{CPV}$ can be measured.

Another type of undesired $\Bsca$-field arises because the molecules move through the $\Evec_\mathrm{MI}$-field with finite velocity $\vec{v} =  v\hat{x}$. They therefore experience a motional magnetic field,
\begin{equation}
    \bm{\mathcal{B}}_\mathrm{mot} = \vec{v}\times\frac{\bm{\mathcal{E}}}{c^2}.
\end{equation}
$\bm{\mathcal{B}}_\mathrm{mot}$ is always perpendicular to both $\vec{v}$ and $\bm{\mathcal{E}}$, i.e., nominally in the $\hat{y}$ direction. If there is any static magnetic field with a nonzero $y$-component, the total magnetic field magnitude will be $\Bsca_\text{tot} = \sqrt{\Bsca_\text{int}^2 + (\Bsca_\mathrm{mot} +\Bsca_\mathrm{st})^2}$.  This means that $\Bsca_\mathrm{tot}$ will change in magnitude when $\Evec_\mathrm{MI}$ is reversed. Since the Zeeman splitting between spin up and down is proportional to $\Bsca_\text{tot}$, this leads to a frequency shift under $\Evec_\mathrm{MI}$ reversal. With our experimental parameters, the resulting shift is approximately $\Bsca_{\mathrm{st},y} \times 50\,$nHz/$\mu$G. Assuming we reach our target level of residual magnetic field, $\Bsca_\mathrm{st} < 10\,\mu$G, a shift of $0.5~\mu$Hz is expected. However, this shift is strongly (but not completely) suppressed due to the $M$ modulation, because the Zeeman shift due to $\Bvec_\mathrm{mot}$ is nearly, but not identically, equal for transitions ej and hk. The difference in the motional-field induced shift between the two transitions is $\approx\! \Bsca_{\mathrm{st},y} \times 5.6\,\mathrm{nHz}/\mu$G. For a field $\Bsca_{\mathrm{st},y} = 10\,\mu$G, this is roughly the same as our anticipated statistical sensitivity.  However, as described in Sec.\ \ref{Sec:InternalComagnetometry}, it should be possible to isolate any residual contribution from the motional field shift by employing co-magnetometry in \CENTREX. 

\subsubsection{Other known sources of systematic errors}
We have considered several other known sources of systematic errors that have been discussed in literature on searches for $T$-violation in TlF.  For example, shifts due to the Millman effect \cite{millman1939determination} (caused by misalignment of the NMR RF field coils) reverse with $B$ and $M$, and hence are suppressed only by $E$ modulation. However, with good construction techniques the residual effects appear likely to be smaller than our anticipated sensitivity.
Furthermore, the Millman effect can be quantified experimentally (see Sec. \ref{Sec:InternalComagnetometry} and Ref. \cite{cho1991search}).
Similarly, we have considered the effect of undesired phase offsets between the two RF coils, and also found the residual effects to be small compared to our anticipated sensitivity.

\subsection{Internal co-magnetometry in \\ \CENTREX\ }
\label{Sec:InternalComagnetometry}

Because the risk of systematic errors from stray magnetic fields is substantial, many of the latest generation of EDM searches have employed co-magnetometers, i.e., other physical systems used to measure magnetic fields co-located with the EDM-sensitive system in both space and time.  Some experiments have used different species nominally sharing the same volume \cite{abel2020measurement,regan2002new}.  Others have used different internal states of the EDM-sensitive system, which have different sensitivity to the EDM and/or to magnetic fields \cite{EckelDeMillePbO_2013,andreev_improved_2018}.  This ``internal co\hyph magnetometer'' approach~\cite{DeMille2001Search} has the advantage of guaranteed spatial overlap between the two systems, and reduced experimental complexity.

We believe it will be possible to use different internal states of TlF to act as a type of generalized internal co-magnetometer. As we have discussed, the apparently natural choices of internal states to use for the $^{205}$Tl NSM search are those where the $^{205}$Tl spin flips, but all other quantum numbers remain (nominally) the same.  This corresponds to the pairs e$\leftrightarrow$j and  h$\leftrightarrow$k assumed throughout our discussion.  However, it is entirely viable to instead employ pairs of states where only the $^{19}$F spin flips, i.e., the pairs e$\leftrightarrow$g and f$\leftrightarrow$h.  As shown in Table~\ref{tab:freq_shifts}, these pairs of states are 2-3 times more sensitive to magnetic field effects than the usual pairs. However, they have negligible sensitivity to $T$-violating effects, since the $^{19}$F nucleus has small $Z$ and $A$.  Hence, these pairs of states can act as a classic co-magnetometer.  The experimental configuration remains nearly unchanged from that used for NSM detection; the primary change is that a significantly lower NMR resonance frequency, $f_0^\prime = 10.6\kHz$, is needed.  We see no impediments to using these pairs of states, which will provide a novel diagnostic for systematic errors and stray fields in \CENTREX. We are still designing state preparation and readout protocols that will enable use of these pairs of states.

Even more potentially useful could be to employ the pairs of states f$\leftrightarrow$k and g$\leftrightarrow$j.  In these transitions, \textit{both} nuclear spins flip simultaneously. Measurements with these pairs are nearly 2 times less sensitive to magnetic fields from leakage currents and residual shield magnetization than the original pair and more than an order of magnitude less sensitive to $\Esca$-induced Zeeman shifts, but have have enhanced sensitivity to motional field shifts.  Hence, making measurements with these pairs as well as both single spin-flip pairs will provide a wealth of information to disentangle contributions from the most important systematic error contributions we are now aware of.  Employing these double spin-flip transitions will require an additional NMR RF coil to produce fields along $\hat{z}$. Here, because of the small transition dipole matrix element, the RF field magnitude will need to be roughly 10 times larger than for the other pairs.  We are currently investigating the feasibility of using these states in \CENTREX.

\section{Conclusion}
As described in section~\ref{sec:sensitivity_and_systematics}, we anticipate a statistical sensitivity to the CPV-induced energy ($\Delta_\mathrm{CPV}$) of $\delta \Delta_\mathrm{CPV} \approx  50~$nHz.  This would correspond to a roughly 2500-fold improvement over the previous best measurements of the $^{205}$Tl NSM.
Taking into account the calculated relation between the NSM and underlying parameters of fundamental physics, this would in many cases correspond to a significantly improved sensitivity over the current best limits.  For example, this would be sensitive to values of the QCD CPV parameter $\bar{\theta} \gtrsim 1 \times 10^{-12}$, a factor of $\approx 100$ smaller than current bounds \cite{abel2020measurement,graner2016reduced}, and to a proton EDM of $d_p \gtrsim 6\times 10^{-27}\,e$cm, a factor of $\approx 30$ smaller than the current best limit \cite{graner2016reduced}.

Currently, measurements and optimization of the rotational cooling efficiency are underway. Once this is completed, the SPA region and then the EQL region will be attached to the beamline for testing and optimization. The Main Interaction region is under construction. The remaining regions, SPB, SPC, and FD, are under design. Once the entire apparatus is assembled and tested, we will commence measurements, with the goal to reach the target sensitivity $\delta \Delta_\mathrm{CPV} \lesssim 50\,$nHz.

Subsequent generations of \CENTREX\, with considerable further improvements in sensitivity, also are anticipated.  For example, we plan to implement transverse laser cooling to collimate the TlF beam \cite{norrgard2017hyperfine, hunter2012prospects}, and a continuous cryogenic buffer gas beam source~\cite{Patterson_2007, Patterson_2009, Patterson_2015, PhysRevA.97.032704} loaded by a thermal TlF beam.  Preliminary estimates indicate that these improvements could increase the detected number of molecules by a factor of 30-100. In the further future, it may also be possible to slow, cool, and optically trap the TlF molecules.  This could dramatically increase the interaction time per molecule, though it remains to be seen what fraction of molecules can be captured in this way. In any case, the \CENTREX\ approach has the potential to yield substantially improved sensitivity to flavor-neutral CPV physics in the hadronic sector.

\CENTREX\ may also be used to search for axions, either measuring the oscillating Schiff moment produced by the interaction with an axion dark matter particle~\cite{PhysRevD.89.043522} or searching for virtual axions mediating CP-violation and producing a Schiff moment in the Tl nucleus~\cite{PhysRevLett.120.013202, PhysRevD.98.035048}.

We thank L.R.\ Hunter and N.\ Clayburn for many helpful discussions, and for sharing preliminary data on optical cycling in TlF. 

We are grateful for support from the John Templeton Foundation, the Heising-Simons Foundation, a NIST Precision Measurement Grant, and NSF-MRI grants PHY-1827906, PHY-1827964, and PHY-1828097, and the US DOE Office of Nuclear Physics.

\appendix
\section{State Evolution and Loss Between Regions in CeNTREX}
\label{sec:non_adiabatic_transitions}

%

In this Appendix, we explain in some detail how quantum states evolve as they move between the different regions of the \CENTREX\ beamline.  Our discussion centers on the mechanisms that lead to undesired population transfer, and their likely magnitude in \CENTREX.

\begin{figure}
	\centering
	\includegraphics[width=0.48\textwidth,unit=1mm]{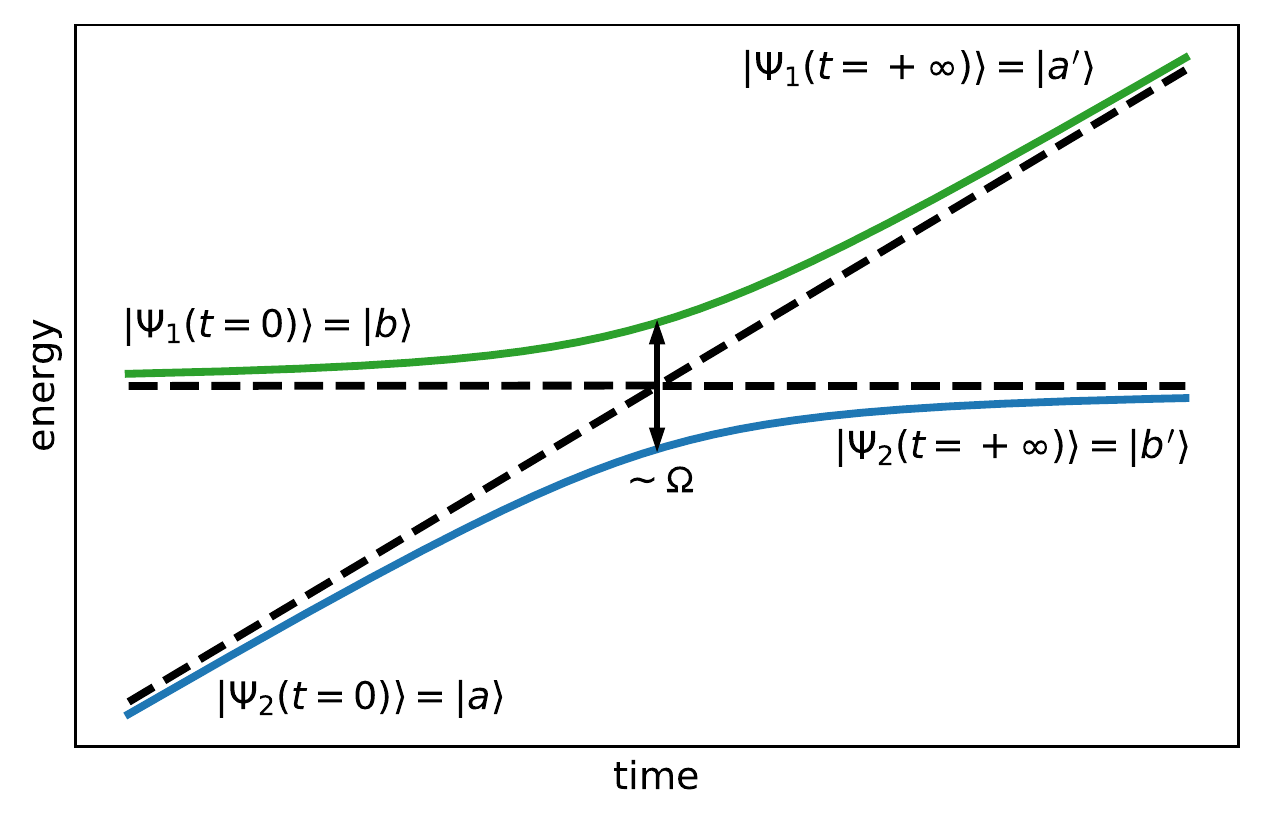}
	\caption{Energy level diagram for a two-level system with an avoided crossing. Black dashed lines show the energies when $\mathcal{H}_\text{I} = 0$; solid lines show energies in the presence of nonzero coupling strength $\hbar\Omega$. Adiabatic (diabatic) evolution corresponds to initial population in one state remaining on the solid (dashed) lines as the system evolves through the avoided crossing.}
	\label{fig:LZ_energy_diagram}
\end{figure}

Unwanted state transfer is most likely to occur when the desired level undergoes an avoided crossing with an undesired level. Such avoided crossings occur in \CENTREX\ when a pair of states are coupled by one mechanism (e.g., hyperfine or Zeeman interactions) while their energy varies due to a separate mechanism (e.g., Stark shifts in varying $\Evec$-fields). 
A qualitative understanding of when transitions occur at a level crossing can be found via generalization of the Landau-Zener model \cite{wittig_landauzener_2005}.  We consider cases where the system begins in the pure state $\ket{a}$, and the time-varying energy splitting $\Delta(t)$ between $\ket{a}$ and the other state, $\ket{b}$, goes through 0. Here $\Delta(t)$ refers to the energy splitting when neglecting terms in the Hamiltonian that couple these states, $\mathcal{H}_\text{I}$. The nonzero coupling between the two states, with strength $\hbar\Omega = \bra{a}\mathcal{H}_\text{I}\ket{b} $, leads to an avoided crossing (Fig.~\ref{fig:LZ_energy_diagram}).

Since the character (i.e., good quantum numbers) of each state can be markedly different on either side of an avoided crossing, we label the upper (lower) state after the crossing as $\ket{a'}$ ($\ket{b'}$).  
The state of the system after the avoided crossing is governed by the parameter $\Gamma = \Omega^2/(d\Delta/dt)$. The probability to end in $\ket{b'}$ is large, i.e., the evolution is adiabatic, when $\Gamma \gg 1$. Conversely, the probability to end in $\ket{a'}$ is large, i.e.\ the evolution is sudden, when $\Gamma \ll 1$. In the intermediate range, when $\Gamma\!\sim\! 1$, the final state is generally a superposition of $\ket{a'}$ and $\ket{b'}$, with relative amplitudes that depend critically on the details of the system.

\CENTREX\ is designed so that molecular states evolve either adiabatically, $\Gamma \gg 1$, or suddenly, $\Gamma \ll 1$, through avoided crossings that occur when traversing between functional regions. Thus, the state before and after any such traversal should be deterministically pure. Throughout the experiment, the local $\Evec$-field is always sufficiently large to define a local quantization axis $\hat{\zeta}$, whose direction changes continuously along the molecular trajectory.  In the frame that is co-moving with the molecules, couplings between desired and undesired states arise from hyperfine interactions, Zeeman interactions, or changes in $\Evec$-field direction.  

Earlier experiments using $^{205}$TlF to search for $T$\hyph violation noted severe problems with deterministic state transfer when molecules move from regions of low $\Esca$ to high $\Esca$. This is likely to occur because of the high density of avoided crossings in this transition between regimes \cite{cho1991search,wilkening1984search}.  Hence, throughout the \CENTREX\ apparatus we ensure that $\Esca > 50 \Vcm$, such that only transitions between mid- and high-field regimes are relevant. There are two classes of transition regions where deterministic evolution of pure states is nontrivial in \CENTREX. The first class refers to transitions between the electrostatic lens and State Preparation regions A and B; the second class refers to transitions between the Main Interaction region and State Preparation regions B and C. We discuss each in some detail here.

The $\Esca$-fields in the SPA and SPB regions ($\sim\!100 \Vcm$ maximum) and in the transitions between these and the EQL region ($\sim\!50 \Vcm$ minimum) lie in the mid-field regime; in the EQL region, $\Esca \sim \hbox{10--30} \kVcm$ is in the high-field regime (see Sec.~\ref{sec:TlF_in_E_fields}). In the transition between mid- and high-field regimes, several subtle but important effects arise due to the coupling between molecular rotation $\vec{J}$ and nuclear spins $\vec{I}_1$ and $\vec{I}_2$ (described by the terms proportional to $c_1 $ and $c_2$ in Eq.~\ref{eq:hyperfine_hamiltonian}). 
First, in the mid-field regime (see Sec. \ref{sec:TlF_in_E_fields}), the molecular eigenstates are only nominally described by the mid-field basis states $\ket{J,m_J}\ket{\It,m_{\It}}$ (for $m_J=0$). This means that molecules nominally prepared in the desired state $ \ket{J=2, m_J = 0}\ket{\It=0,m_{\It}=0} $, in the SPA region, are actually prepared in an eigenstate $\ket{\psi_{\rm SPA}}$ that has a small admixture of states with $ m_J = \pm 1 $.
For example, when $\Esca \sim 50 \Vcm$, 
we find
\begin{equation}
    \begin{split}
    	\ket{\psi_{\rm SPA}} \approx{} & \ket{J=2,m_J=0}\ket{\It=0,m_{\It}=0} \label{eq:SPA_ket}\\
        & {}+\eta \ket{J = 2, m_J = 1}\ket{\It=1,m_{\It}=-1} \\
    	 & {}-\eta \ket{J = 2, m_J = -1}\ket{\It=1,m_{\It}=+1},
    \end{split}
\end{equation}
where the mixing coefficient $\eta$ is determined by the strength of the hyperfine interaction compared to the Stark shift between states with different $m_J$: $\eta \approx \hbar\Omega_{\rm hf}/\Delta E_{\rm S} \sim 0.1$.  Here, the nonzero value of $\eta$ arises from the spin-rotation terms in $\mathcal{H}_\text{I}$, which couple states with $\Delta m_J = \pm 1 = -\Delta m_{\It}$. By contrast, in the high-field regime of the EQL, states with different values of $m_J$ are very distant. Hence, here the true eigenstates $\ket{\psi_{\rm EQL}}$, corresponding to the desired states $\ket{\widetilde{J},m_J=0}\ket{\It=0,m_{\It}=0}$, have negligibly small admixtures of states with $m_J\neq 0$ or $\It\neq 0$ (i.e., $\eta \lll 1$).

\begin{figure}
	\centering
	\includegraphics[width=0.48\textwidth,unit=1mm]{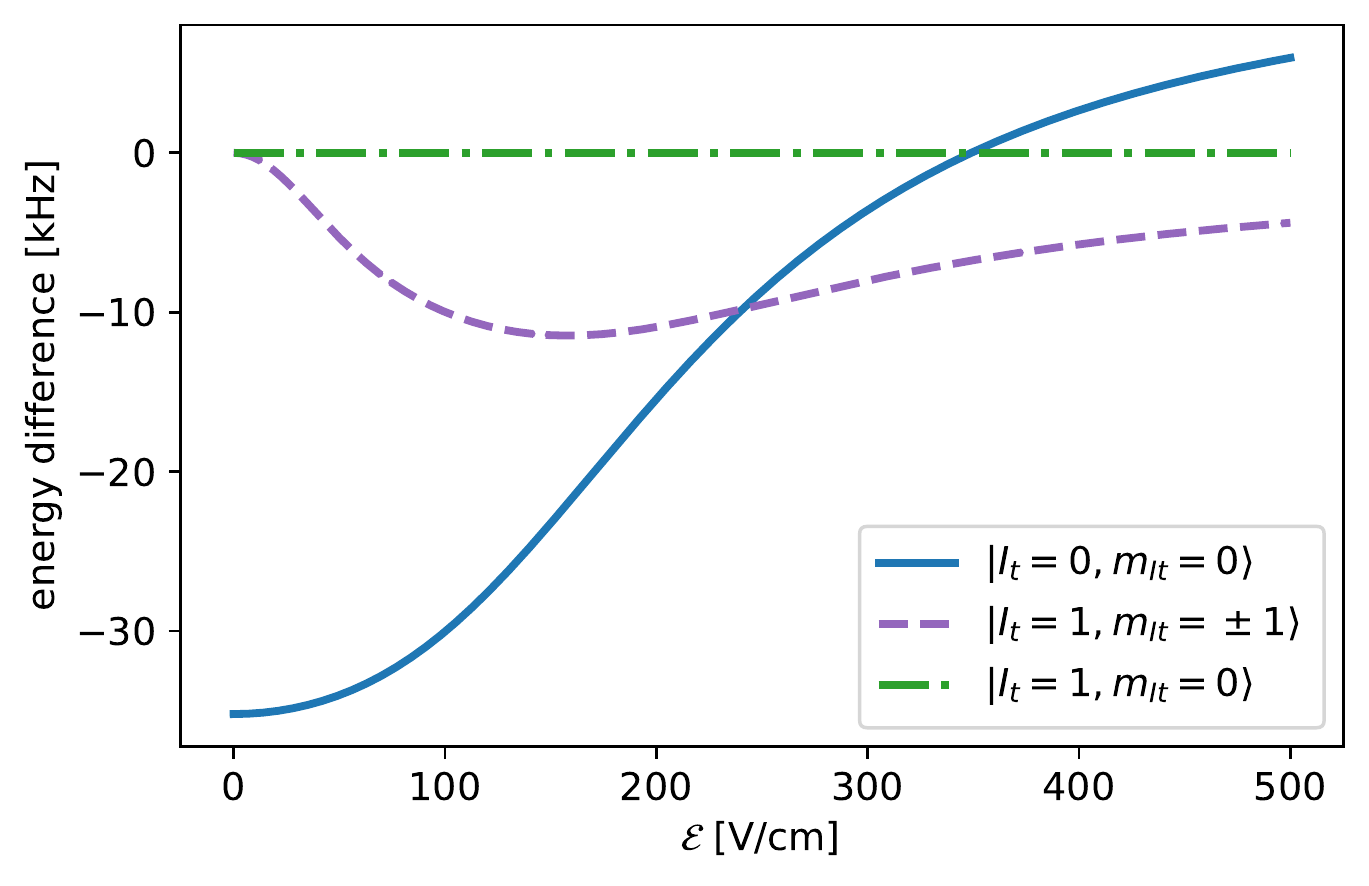}

	\caption{Relative energies of the nuclear spin states within the $J=2, m_J = 0$ manifold, versus $\Esca$-field magnitude (with $\Bsca = 0$). The energy of the $\ket{\It=1, m_{\It} = 0}$ state is defined as the reference energy at every value of $\Esca$. Level crossings occur at $\sim240$ and $350 \Vcm$. If the $\Bvec$-field is non-zero or the $\Evec$-field is rotating, the different spin states will be coupled and the crossings will be avoided.}
	\label{fig:spin_state_crossings}
\end{figure}

Second, as a function of $\Esca$ in the transition from mid- to high-field regimes, the (nominal) $\ket{J=2,m_J=0}$ eigenstates with different spin content undergo a few level crossings (Fig.~\ref{fig:spin_state_crossings}).
This occurs as the spins fully decouple from rotation in any $m_J=0$ rotational state.

Finally, in the  mid-field regime, rotation of the $\Evec$-field can cause transitions between states with (nominally) different nuclear spin configurations. The source of these transitions can be understood qualitatively: in the mid-to-high field regime, $ \vec{J} $ is strongly coupled to $\Evec$ and must reorient appropriately as the electric field rotates. If the rotation of $ \vec{J} $ is too fast for the coupled nuclear spins to follow, their orientation with respect to the quantization axis provided by $ \Evec $ may change so that the spins end up in a different state relative to the local $\hat{\zeta}$-axis.

To describe couplings induced by $\Evec$-field rotation, we follow the approach of Wall et al.\ \cite{wall_nonadiabatic_2010}. We write the Hamiltonian in a co-moving frame with axes $(\xi,\upsilon,\zeta)$, where $\Evec$ defines the local direction of the $\zeta$-axis at all points along the molecular trajectory. In this frame, $\hat{\zeta}$ may point in any direction relative to a set of laboratory-fixed axes, and its direction rotates continuously as the molecules move along their path in the lab. Consider what happens when a molecule moves from the low field in the SPA region to the high field in the EQL region. In the SPA region, $\Evec$ is parallel to the average molecular beam direction, $\hat{Z}$. The quadrupole field is always in the $X-Y$ plane in this frame; here we consider a particular molecular trajectory such that the direction of $\Evec$ inside the lens is along $\hat{X}$. (Analogous arguments hold for other trajectories.)

\begin{figure}
	\centering
	\includegraphics[width=0.48\textwidth,unit=1mm]{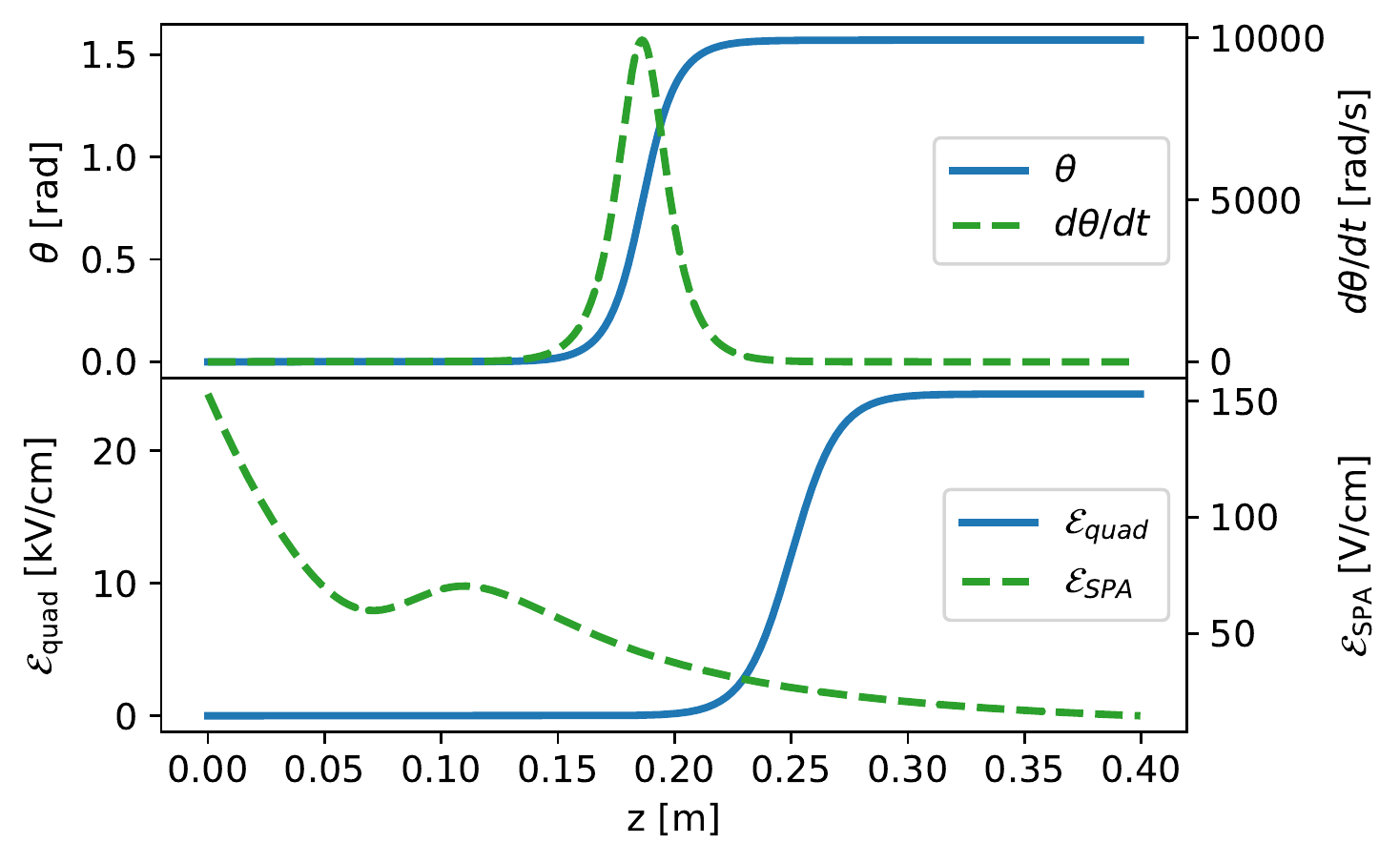}
	\caption{\textbf{Top:} Angle $\theta$ between the $\Evec$-field direction $\hat{\zeta}$ and the lab frame $Z$-axis, and its rate of change over time $d\theta/dt$, for a molecule with typical velocity $\vec{v} = v_Z\hat{Z}$ and $v_Z = 184 \;\mathrm{m/s}$, vs $Z$-position. \textbf{Bottom:} The magnitudes of the $\Esca$-fields in the transition region between the SPL and EQL regions, as a function of $Z$-position. The field due to the SPA region electrodes, $\Evec_\text{SPA}$, is along the length of the apparatus ($Z$); the direction of the field due to the lens electrodes, $\Evec_\text{EQL}$, is taken to be along $X$ in the lab frame. }
	\label{fig:rotating_electric_field}
\end{figure}

To keep the $\Evec$-field along $\hat{\zeta}$, we rotate co-moving coordinate system about the laboratory $Y$-axis, by an angle $ \theta(t) = \arctan\left(\frac{\Esca_{\rm EQL}(t)}{\Esca_{\rm SPA}(t)}\right)$. Here, $ \Esca_{\rm EQL}(t) $ is the magnitude of the transverse field due to the EQL, and $ \Esca_{\rm SPA}(t) $ that of the axial field due to the ring electrodes in the SPA region. As the molecule moves between the regions, it will see the electric field rotate as $\Esca_{\rm SPA}$ diminishes and $\Esca_{\rm EQL}$ increases (see Figure~\ref{fig:rotating_electric_field}). The unitary rotation matrix that takes us from the lab frame to the co-moving frame is given by 
\begin{equation}
	U_\text{R}(\theta) = \exp(-i\theta (t) \vec{F}\cdot\hat{Y}),
\end{equation}
where $\vec{F} = \vec{J} + \vec{I}_1 + \vec{I}_2$ is the total angular momentum. The time-evolution in this rotated frame is given by
\begin{equation}
i\hbar\frac{d\ket\psi_\text{R}}{dt} = \mathcal{H}_\text{eff}\ket\psi_\text{R} = 
\left(U_\text{R}^\dagger \mathcal{H} U_\text{R} - i\hbar U_\text{R}^\dagger\frac{dU_\text{R}}{dt}\right)\ket{\psi_\text{R}},
\end{equation}
where $\mathcal{H}$ is the Hamiltonian in the lab frame and $\ket{\psi_\text{R}}$ is the state vector in the rotated frame. The term $U_\text{R}^\dagger \mathcal{H} U_\text{R}$ is the usual Hamiltonian for TlF with a time-varying but non-rotating $\Evec$-field along $\hat{\zeta}$. The other term, $U_\text{R}^\dagger\frac{dU_\text{R}}{dt}$, contains the effects due to the rotation of $\Evec$. This term can be written as
\begin{equation}
\mathcal{H}_\text{int}^\text{(eff)} = - i\hbar U_\text{R}^\dagger\frac{dU_\text{R}}{dt} =  -\hbar\left(J_Y+I_{1,Y}+I_{2,Y}\right)\frac{d\theta}{dt}.
\end{equation}

Due to the spin-rotation interaction, the matrix elements of $J_Y$ between $\ket{\psi_{\rm SPA}}$ and the undesired spin triplet states, nominally $\ket{J=2, m_J = 0}\ket{\It=1,m_{\It}=\mp 1}$, are non-zero; their magnitude is $\Omega \sim \eta \frac{d\theta}{dt}$, where $\eta$ is the mixing coefficient from Eq.~\ref{eq:SPA_ket}. Due to this off-diagonal coupling, the level crossings in Fig.~\ref{fig:spin_state_crossings} become avoided crossings. Hence, fully adiabatic evolution here would result in our desired $\It=0$ state evolving into an undesired $\It=1$ state, as shown in Fig.~\ref{fig:LZ_energy_diagram}. Instead, here we want the state evolution to be sudden/fully diabatic to maintain $\It=0$. To avoid population loss, we thus require $ d\Delta/dt \gg \eta^2 \dot{\theta}^2 $. This condition only needs to be fulfilled when the coupled levels are close in energy, i.e.\ when $\Esca \approx 200-400 \Vcm$ (see Fig.~\ref{fig:spin_state_crossings}). This is achieved in practice by allowing $\Esca_{\rm SPA}$ to decay to $ \approx 50 \Vcm $ before $\Esca_{\rm EQL}$ starts to rapidly increase. The $\Evec$-field is then almost entirely in the transverse direction, i.e., not rotating quickly, by the time the level-crossing occurs (Fig.~\ref{fig:rotating_electric_field}).

We note in passing that the axial-to-transverse field configuration in \CENTREX\ has not been used in previous $^{205}$TlF experiments.  If only transverse fields are used, inevitably some large fraction of molecular trajectories travel through a position where $\Esca=0$ and undesired transitions are strong.  Our approach mimics that of Ref.~\cite{WoodWiemanParity1997}, but using $\Evec$ rather than $\Bvec$ as the quantizing field.

Magnetic fields can also couple the desired $\It=0$, $m_{\It} = 0$ state to $\It=1$ states with $m_{\It}=0$ ($m_{\It} = \pm 1$) when $\Bsca_Z \neq 0$ ($\Bsca_{X,Y} \neq 0$).  To reduce this effect, in \CENTREX\ we will apply shim coils to cancel typical ambient lab fields in the regions of transition into and out of the EQL region. With cancellation by a factor of $\gtrsim 10$, such that $\Bsca \lesssim 0.05\mG$, the Zeeman coupling strengths are in the range $\Omega_Z\lesssim 0.1\kHz$. When the condition needed to avoid transitions from $\Evec$-rotation given above are satisfied, the rate of change of the level splittings as molecules enter the very strong $\Esca$-field of the lens, $d\Delta/dt$, is sufficiently large compared to $\Omega_Z$ such that the evolution is fully diabatic. Numerical simulations indicate a loss of $<1\%$ along any molecular trajectory. Hence, under these conditions the quantum numbers $\It=0,m_{\It}=0$ are preserved as molecules enter and exit the EQL region.

The transition from the EQL region to SPB region is mostly similar to the transition from the SPA region to the EQL region. The primary difference is the requirement in the SPB region to have a uniform electric field $\Evec_{\rm SPB} = \Esca_{\rm SPB} \hat{z}$, along with a substantial magnetic field, $\Bvec_{\rm SPB} = \Bsca_{\rm SPB}\hat{z}$, where $\Bsca_{\rm SPB} \approx \hbox{10--20}\gauss$. Here, we are describing fields in the $(x,y,z)$ ``interaction region'' coordinate system. These fields can be reached by first diabatically rotating from the large transverse lens field $\Evec_{\rm EQL}$ to a weak axial field $\Evec \parallel \hat{Z}$, then adiabatically rotating into the uniform transverse field $\Evec_{\rm SPB} \parallel \hat{z}$. Throughout the second $\Evec$ rotation, $\Esca$ remains in the range $\hbox{50--100} \Vcm$ while $\Bsca_z$ slowly rises, from its initial value of (nominally) zero to $\Bsca_{\rm SPB}$. Though the details remain to be worked out, this scheme should ensure deterministic population of the desired state in the EQL--SPB transition.

The last class of traversals in \CENTREX\ occurs between the SPB and MI regions (or, similarly aside from the reversed sequence, the MI and SPC regions).  Here, the $\Bsca$-field must transform from $\Bsca_{\rm SPB}$ to zero, and the $\Evec$-field can remain in the same $\hat{z}$ direction but must make the transition from low- to high-field regimes.  This can again be accomplished by adiabatically ramping $\Bsca$ to zero while maintaining $\Esca \approx \hbox{50--100} \Vcm$ along $\hat{z}$.  Then, a sudden rise to $\Esca \gg 500 \Vcm$ will maintain the spin quantum numbers for molecules coming into/out of the MI region. During these traversals, there is a possibility of undesired transitions between different nuclear spin states, if the $\Bvec$-field is not fully parallel to the $\Evec$-field. Whether or not these transitions are likely to pose a problem, and if so how to mitigate them, is currently being investigated. We note that in prior experiments with TlF \cite{wilkening1984search,cho1991tight}, this issue was appreciated but not fully under control.
                       
While keeping track of state evolution across level crossings may appear daunting, it is analogous to even more complex schemes that have been applied efficiently in other molecular systems \cite{langheckerdenschlag2008}. We believe that our detailed understanding of and control over these issues will be necessary to understand and minimize systematic errors in \CENTREX.

{

	\small
	\bibliographystyle{unsrtnat}
	\bibliography{references}
}

\end{document}